\documentclass[12pt]{amsart}

\usepackage{cite,amssymb,color}

\usepackage{graphicx}

\newtheorem{theorem}{Theorem}[section]
\newtheorem{lemma}[theorem]{Lemma}

\theoremstyle{definition}
\newtheorem{definition}[theorem]{Definition}

\theoremstyle{theorem}
\newtheorem{meta}[theorem]{Meta-Theorem}

\theoremstyle{corollary}
\newtheorem{corollary}[theorem]{Corollary}

\theoremstyle{conjecture}
\newtheorem{conjecture}[theorem]{Conjecture}

\theoremstyle{proposition}
\newtheorem{proposition}[theorem]{Proposition}

\theoremstyle{remark}
\newtheorem{remark}[theorem]{Remark}

\numberwithin{equation}{section}
\newcommand{\R}{\mathbb{R}}

\newcommand{\Z}{\mathbb{Z}}
\newcommand{\C}{\mathbb{C}}
\newcommand{\T}{\mathbb{T}}

\newcommand{\B}{\mathbf{B}}
\newcommand{\F}{\mathcal{F}}

\newcommand{\Q}{\mathbb{Q}}
\newcommand{\G}{\Gamma}
\newcommand{\g}{\gamma}
\newcommand{\U}{\mathcal{U}_\G}

\newcommand{\bd}{\begin{definition}}
\newcommand{\ed}{\end{definition}}
\newcommand{\bt}{\begin{theorem}}
\newcommand{\et}{\end{theorem}}
\newcommand{\bl}{\begin{lemma}}
\newcommand{\el}{\end{lemma}}
\newcommand{\bc}{\begin{corollary}}
\newcommand{\ec}{\end{corollary}}
\newcommand{\bcon}{\begin{conjecture}}
\newcommand{\econ}{\end{conjecture}}
\newcommand{\br}{\begin{remark}}
\newcommand{\er}{\end{remark}}
\newcommand{\bp}{\begin{proposition}}
\newcommand{\ep}{\end{proposition}}
\newcommand{\be}{\begin{equation}}
\newcommand{\ee}{\end{equation}}
\newcommand{\bef}{\begin{figure}}
\newcommand{\eef}{\end{figure}}
\newcommand{\bea}{\begin{eqarray}}
\newcommand{\eea}{\end{eqarray}}
\newcommand{\ba}{\begin{array}}
\newcommand{\ea}{\end{array}}
\def\H{\mathcal{H}}
\def\B{\mathcal{B}}
\DeclareMathOperator{\Tr}{Tr}

\newcount\theTime
\newcount\theHour
\newcount\theMinute
\newcount\theMinuteTens
\newcount\theScratch
\theTime=\number\time
\theHour=\theTime
\divide\theHour by 60
\theScratch=\theHour
\multiply\theScratch by 60
\theMinute=\theTime
\advance\theMinute by -\theScratch
\theMinuteTens=\theMinute
\divide\theMinuteTens by 10
\theScratch=\theMinuteTens
\multiply\theScratch by 10
\advance\theMinute by -\theScratch

\def\today{{\number\day\space
 \ifcase\month\or
  January\or February\or March\or April\or May\or June\or
  July\or August\or September\or October\or November\or December\fi
 \space\number\year}}

\begin{document}

\title{An overview of periodic elliptic operators}


\author{Peter Kuchment}
\address{Mathematics Department,Texas A\&M University, College Station, TX 77843-3368}
\email{kuchment@math.tamu.edu}
\thanks{This article contains an extended exposition of the lectures given at Isaac Newton Institute for Mathematical Sciences in January 2015.
The work was also partially supported by the NSF grant DMS \# 1517938. The author expresses his gratitude to the INI and NSF for the support.}





\dedicatory{Dedicated to the memory of mathematicians and dear friends\\M.~Birman, L.~Ehrenpreis, S.~Krein, and M.~Novitskii}

\begin{abstract}
The article surveys the main topics, techniques, and results of the theory of periodic operators arising in mathematical physics and other areas. Close attention is paid to studying analytic properties of Bloch and Fermi varieties, which influence significantly most spectral features of such operators.

The approaches described are applicable not only to the standard model example of Schr\"odinger operator with periodic electric potential $-\Delta+V(x)$, but to a wide variety of elliptic periodic equations and systems, equations on graphs, $\overline{\partial}$-operator, and other operators on abelian coverings of compact bases.

Important applications are mentioned. However, due to the size restrictions, they are not dealt with in details.
\end{abstract}

\maketitle
\tableofcontents
\section*{Introduction}
Elliptic PDEs with periodic coefficients, notably the stationary\\ Schr\"odinger operator $-\Delta +V(x)$ with a periodic potential\footnote{The potential is usually, but not always, is considered to be real.} $V$, have been intensively studied in physics and
mathematics literature for close to a century, due to their crucial role in the solid state theory, as well as in other areas (e.g., see books and surveys \cite{Brillouin_periodic,BrilExtended,Bethe_elektronen,AshcroftMermin_solid,Ziman_solids,Ziman_Fermi,Kittel_solidstate,ReedSimon_v14}). In spite of that, some important questions
have remained unresolved. The interest in periodic elliptic (and sometimes parabolic) equations and systems received a strong boost in the last decades, due to
new applications in areas such as photonic crystals and other metamaterials \cite{Kuch_pbg}, fluid
dynamics \cite{Yudovich_book,Conca_book}, carbon nanostructures \cite{KatsGraph}, inverse scattering method of solving integrable systems \cite{Novikov_viniti,Nov_sol84}, and lately topological
insulators \cite{BernTopIns}. When the author was preparing his introductory lectures at the Newton Institute in January 2015 \cite{KucINI15}, he discovered that there was no comprehensive source devoted to the
theory of periodic elliptic PDEs and their applications. There are several books and surveys completely or partially devoted to this topic (e.g.,\cite{ShubinBerezin_schrod,Veliev_book,AllairConcaVann,kuch_floquet1982,Karp_LNM,Kuc_floquet,ReedSimon_v14,Eastham_periodic,Brown_per_book,TitchV2,Skr_psim85,Gieseker_fermi,Ger_bsmf90,BerKuc_book}), but none of them collects most of the useful techniques (e.g., the analytic
geometry of Bloch and Fermi varieties) and is up to date. Thus, the idea was born to expand the lectures to this survey that would contain the main techniques and results of the spectral theory of
periodic operators arising in mathematical physics and other areas. These are  applicable not only to the standard model example of Schr\"odinger operator with periodic electric potential
$-\Delta+V(x)$, but to a wide variety of elliptic periodic equations and systems, equations on graphs, $\overline{\partial}$-operator, and other operators on abelian coverings of compact bases.
Close attention is paid to studying analytic properties of Bloch and Fermi varieties, which influence significantly most properties of such operators.

There are many important applications of what is discussed. However, the author realized at this stage that even a large survey cannot do justice to such applications, and even to many details
of techniques
employed. Thus, several sections (especially toward the end of the text) contain mostly hints and pointers to the literature. The choice has been made according to the author's tastes and might not satisfy everyone. The task of more detailed and comprehensive discussion is postponed
till the third iteration of my lectures, which is a monograph in preparation. The same applies to the bibliography of this survey, which is extensive, but far from being
comprehensive, and many references are given through secondary sources (books and surveys).


\section{$1D$ - a brief sketch}\label{S:1D}

In this section, we survey briefly what is commonly called Floquet (or Floquet-Lyapunov) theory \cite{Floquet,Lyap1896,Lyap1899},
the main tool in studying periodic linear ODEs and systems of ODEs. Its basics can be found in
many ODE textbooks, e.g. in
\cite{CoddCarlson,CoddLevinson,ArnoldGeomODE,ArnoldODE}. There are also nice books and surveys dedicated to
(mostly spectral theory of) the  periodic ODEs of the second order, see e.g. \cite{Arscott,MagnusWinkl_hills,Eastham_periodic,MityaginDzhakov}, \cite[Section 2.8]{ShubinBerezin_schrod}, \cite[Ch. 5]{Simon_Szego}, \cite[Ch. XXI]{TitchV2} and \cite[Section XIII.16]{ReedSimon_v14} (see also \cite[and references therein]{DM09} for the case of singular potentials). I want to attract the reader's attention to the not sufficiently well known amazing treatise \cite{Iakubovich_periodic}, which contains an enormous amount of information concerning periodic ODEs and systems of ODEs, including Hamiltonian and canonical systems, parametric resonance, stability domains, various applications, etc. Here, we will touch upon a few basic things only.

\subsection{Euler's theorem}\label{SS:Euler}
Consider a linear system of ODEs
\begin{equation}\label{E:const}
\frac{dx}{dt}=Cx, t\in\R, x\in\C^n,
\end{equation}
with a constant $n\times n$ matrix $C$. Then all its solutions look as follows:
$$
x(t)=e^{Ct}x_0,
$$
where $x_0$ is the initial value of $x(t)$. One thus obtains the following \textbf{Euler's theorem} \cite{Euler}, which can be found in any ODE textbook:
\bt\label{T:Euler}
All solutions of (\ref{E:const}) are linear combinations of the exponential-polynomial solutions of the form
\be\label{E:exppol}
x(t)=e^{i\lambda t}\sum\limits_{j\in\Z, j\geq 0}p_jt^j,
\ee
where $i\lambda$ is an eigenvalue of $C$ and the sum is finite.

If $C$ has no Jordan blocks, then only $j=0$ is present in (\ref{E:exppol}).
\et

Floquet-Lyapunov theory, sketched below, generalizes this result to the case of systems with periodic coefficients.

\subsection{Floquet-Lyapunov theory}
Let us consider a linear system of ODEs
\begin{equation}\label{E:1Dsystem}
\frac{dx}{dt}=A(t)x, t\in\R, x\in\C^n,
\end{equation}
with a $1$-periodic $n\times n$ matrix function $A(t)$. As promised before, we will not dwell here on imposing weakest conditions on the entries of the matrix \cite[Ch. II]{Iakubovich_periodic}, assuming that they are continuous. Let $X(t)$ be the $n\times n$ matrix \textbf{fundamental solution} of (\ref{E:1Dsystem}). In other words, if
$$
e_j=(0,\dots,0, \mathop{\underbrace{1}}\limits_{j\mbox{th entry}}, 0, \dots,0)
$$
is the standard basis of $\C^n$ and $x_j(t)$ - the solution of (\ref{E:1Dsystem}) such that $x_j(0)=e_j$, then  $X(t)$ has each $x_j(t)$ as the $j$th column. One thinks of $X(t)$ as the operator that shifts by time $t$ along the trajectories. In particular, when the value of $t$ coincides with the period (i.e., is equal to $1$ under our assumption), we introduce the following

\bd\label{D:monodromy} The matrix
\be\label{E:monodr}
M:=X(1)
 \ee
 of the shift by the period $1$ along the trajectories of the system is said to be the \textbf{monodromy matrix} of the equation (\ref{E:1Dsystem}).
\ed

The main result of the Floquet theory is the following

\begin{theorem}\label{T:floquet}
\textbf{Floquet theorem}\\
There exists a $1$-periodic matrix function $P(t)$ and a constant matrix $C$, such that
\be\label{E:floquet}
X(t)=P(t)e^{Ct}.
\ee
\et

\br
Notice that
\be\label{E:monodr_exp}
M=X(1)=e^C.
\ee
\er

\bc
Any solution of (\ref{E:1Dsystem}) is a linear combination of \textbf{Floquet (or Floquet-Bloch) solutions}
\be\label{E:fl_sol1D}
x(t)=e^{ikt}\sum_{j\in\Z^+} p_j(t)t^j,
\ee
where the sum is finite, coefficients $p_j(t)$
are $1$-periodic, and $e^{ik}$ are the eigenvalues of the monodromy matrix $M$.

When the monodromy matrix does not have Jordan blocks, only the term with $j=0$ is present. In this case,
the solution is $x(t)=e^{ikt}p(t)$ with a $1$-periodic function $p(t)$ and is sometimes called a \textbf{Bloch solution}.
\ec

\bd\label{D:mult_quasim}
The eigenvalues $z:=e^{ik}$ of the monodromy matrix $M$ are called \textbf{Floquet multipliers} and numbers $k$ - \textbf{quasimomenta} or \textbf{crystal momenta} (the latter names coming from solid state physics \cite{AshcroftMermin_solid}).
\ed
\br
One notices that the value of a quasimomentum $k$ is defined only modulo $2\pi\Z$-shifts\footnote{See the discussion of dual lattices and quasimomenta in Section \ref{S:lattices}.}.
\er

In fact, the following, stronger than Theorem \ref{T:floquet}, result holds:
\bt\label{T:lyap}
\textbf{Lyapunov reduction theorem.}
There exists a periodic invertible matrix-function $B(t)$ such that the substitution $x(t)=B(t)y(t)$ reduces the system  (\ref{E:1Dsystem}) to the one with a constant matrix $C$:
$$\frac{dy}{dt}=Cy(t).$$
\et

\subsection{Hill operator}
Studying higher order ODEs with periodic coefficients can be reduced to the case of first order systems. It is, however, useful to consider the special case of the so called \textbf{Hill operator}:
\be\label{E:Hill}
H=-\frac{d^2}{dx^2}+V(x)
\ee
with a ``nice'' real $1$-periodic potential $V$ on $\R$. Again, we will assume the excessive condition of continuity of $V$, although much weaker conditions suffice (see, e.g., \cite{Iakubovich_periodic,ReedSimon_v14,MagnusWinkl_hills,Eastham_periodic,MityaginDzhakov}).
The domain of the operator will be, by definition, the Sobolev space $H^2(\R)$. Defined this way, the operator is self-adjoint\footnote{It is not hard to show that if one defines the operator on the space $C^\infty_0 (\R)$, it will be essentially self-adjoint, with the only self-adjoint extension described above.} \cite{Eastham_periodic}.

\subsubsection{Monodromy, discriminant, and such}
We will be interested in the spectral problem for the Hill operator:
\be\label{E:Hill_spectr}
-\frac{d^2u}{dx^2}+V(x)u=\lambda u.
\ee
In accordance to the general Floquet theory approach, we consider the fundamental system of two (analytic in $\lambda$) solutions $\phi(x,\lambda),\psi(x,\lambda)$:
\be\label{E:HillFundSol}
\phi(0,\lambda)=\psi'_x(0,\lambda)=1, \phi'_x(0,\lambda)=\psi(0,\lambda)=0.
\ee
This enables us to consider the \textbf{monodromy matrix}
\be\label{E:HillMonodr}
M(\lambda):=
\left(
  \begin{array}{cc}
    \phi(1,\lambda) & \psi(1,\lambda) \\
    \phi'_x(1,\lambda)& \psi'_x(1,\lambda) \\
  \end{array}
\right).
\ee
The determinant $\det M(\lambda)$ is the Wronskian of this fundamental system of solutions, evaluated at $t=1$. Since the Wronskian is constant in time and equal to $1$ at $t=0$, we conclude that
\be\label{E:monodr_determ}
\det M(\lambda)=1.
\ee
Thus, since we are interested in the eigenvalues of the monodromy matrix, all the pertinent information is contained in the trace of $M$:
\be\label{E:discr}
\Delta(\lambda):=\Tr M(\lambda)=\phi(1,\lambda)+\psi'_x(1,\lambda),
\ee
which is called  the \textbf{discriminant}, or the \textbf{Lyapunov function} of the Hill operator (\ref{E:Hill}).

Thus, all Floquet multipliers $z$, being the eigenvalues of $M$, can be found from the secular equation
\be\label{E:secular}
z^2 - \Delta(\lambda)z +1=0.
\ee
Its roots provide Floquet multipliers and quasimomenta  for a given $\lambda$:
\be\label{E:quadr_formula}
e^{ik}=z=0.5\left(\Delta(\lambda)\pm \sqrt{\Delta(\lambda)^2-4}\right).
\ee
It is now easy to come up with the following result (a compilation of various statements from standard books, e.g. \cite{Iakubovich_periodic,Eastham_periodic}):
\bt\label{T:role_of_2} \indent \begin{enumerate}
\item If $|\Delta(\lambda)|<2$, then
\begin{itemize}
\item the Floquet multipliers are complex, distinct, of the absolute value $1$, and complex conjugate to each other.
\item In particular, the quasimomentum is real and if $k$ is a quasi-momentum, then $-k$ is\footnote{This observation also follows from the fact that, due to the potential being real and for real $\lambda$, complex conjugate of a solution is a solution as well. In physics terms, this is a reflection of the \textbf{time reversibility} of the dynamical system with the Hill operator as the Hamiltonian.}.
\item All solutions of $Hu=\lambda u$ are bounded.
\item The \textbf{Lyapunov exponent}\footnote{The Lyapunov exponent $r$ characterises the maximal possible rate $e^{rt}$ of the exponential growth of solutions.} is equal to zero.
\end{itemize}
\item If $|\Delta(\lambda)|=2$, then
\begin{itemize}
\item The two Floquet multipliers coincide and are equal to $1$ or $-1$.
\item Quasimomentum is either $k=0$, or $k=\pi\,(\mod 2 \pi)$.
\item All solutions of $Hu=\lambda u$ are bounded or polynomially bounded.
\item The Lyapunov exponent is equal to zero.
\end{itemize}
\item If $|\Delta(\lambda)|>2$, then
\begin{itemize}
\item the Floquet multipliers are real, distinct, and reciprocal to each other. \item Quasimomenta are complex.
    \item Solutions of $Hu=\lambda u$ grow exponentially.
    \item The Lyapunov exponent is positive and equal to $|\Im k|$.
\end{itemize}
\end{enumerate}
\et

Let us look at the case of the free operator (i.e., with zero potential). The straightforward calculation shows then that the free discriminant is $\Delta_0(\lambda)=2\cos(\sqrt{\lambda})$. In particular, it is an entire function of exponential order $1/2$, i.e. $$|\Delta_0(\lambda)|\leq Ce^{C|\lambda|^{1/2}}$$
for some $C>0$.
Its graph looks as shown in Fig. \ref{F:freeDelta}.
\bef[ht!]
\begin{center}
\includegraphics[scale=0.9]{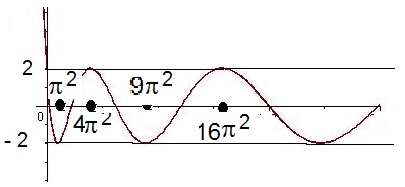}
\caption{The graph of the free discriminant.}\label{F:freeDelta}
\end{center}
\eef
The properties of the discriminant in presence of a periodic potential have also been studied thoroughly (see, e.g., \cite{Eastham_periodic,MagnusWinkl_hills,TitchV2}). As in the free case, it is an entire function of exponential order $1/2$:
$$|\Delta(\lambda)|\leq Ce^{C|\lambda|^{1/2}}.$$
The Figure \ref{F:Delta} gives an idea of its behavior.
\bef[ht!]
\begin{center}
\includegraphics[scale=0.4]{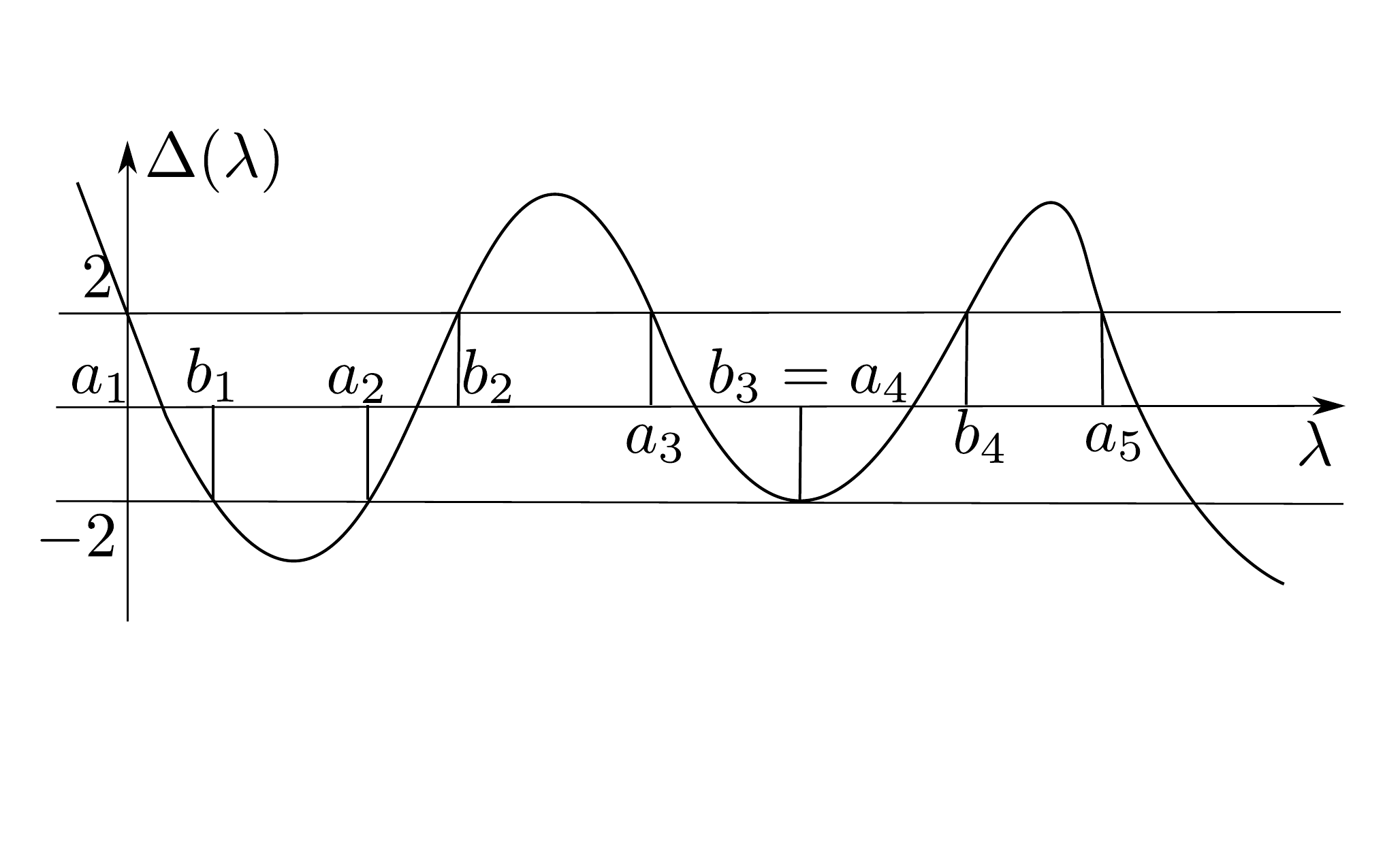}
\caption{Discriminant for a periodic Hill equation.}\label{F:Delta}
\end{center}
\eef

Notice that the extrema of the free discriminant are located exactly on the boundary of the horizontal strip $|\Delta|\leq 2$. They get shifted in the periodic case (without any new ones being created), but cannot move into the interior of this strip (see detailed considerations, e.g., in \cite{Eastham_periodic}). In particular, the pieces of the function $\Delta(\lambda)$ between its consecutive hits of the lines $\Delta=\pm 2$, are \textbf{monotonic}. We will provide a  simple explanation of this monotonicity a little bit later. The reader will see later on that absence of any analog of this monotonicity in higher dimensions is responsible for significant changes in some spectral properties in comparison with $1D$.

\bd The closures of the segments where $|\Delta(\lambda)|< 2$ (shown in \textcolor[rgb]{1.00,0.00,0.00}{red} in the figure below) are called the \textbf{stability zones} or \textbf{spectral bands}. The segments between the bands are called \textbf{instability zones} or \textbf{spectral gaps}. We will denote by $g_n\geq 0$ the length of the $n$th gap, $n=1, 2, \dots$.
\bef[ht!]\label{F:1Dband-gap}
\begin{center}
\includegraphics[scale=0.4]{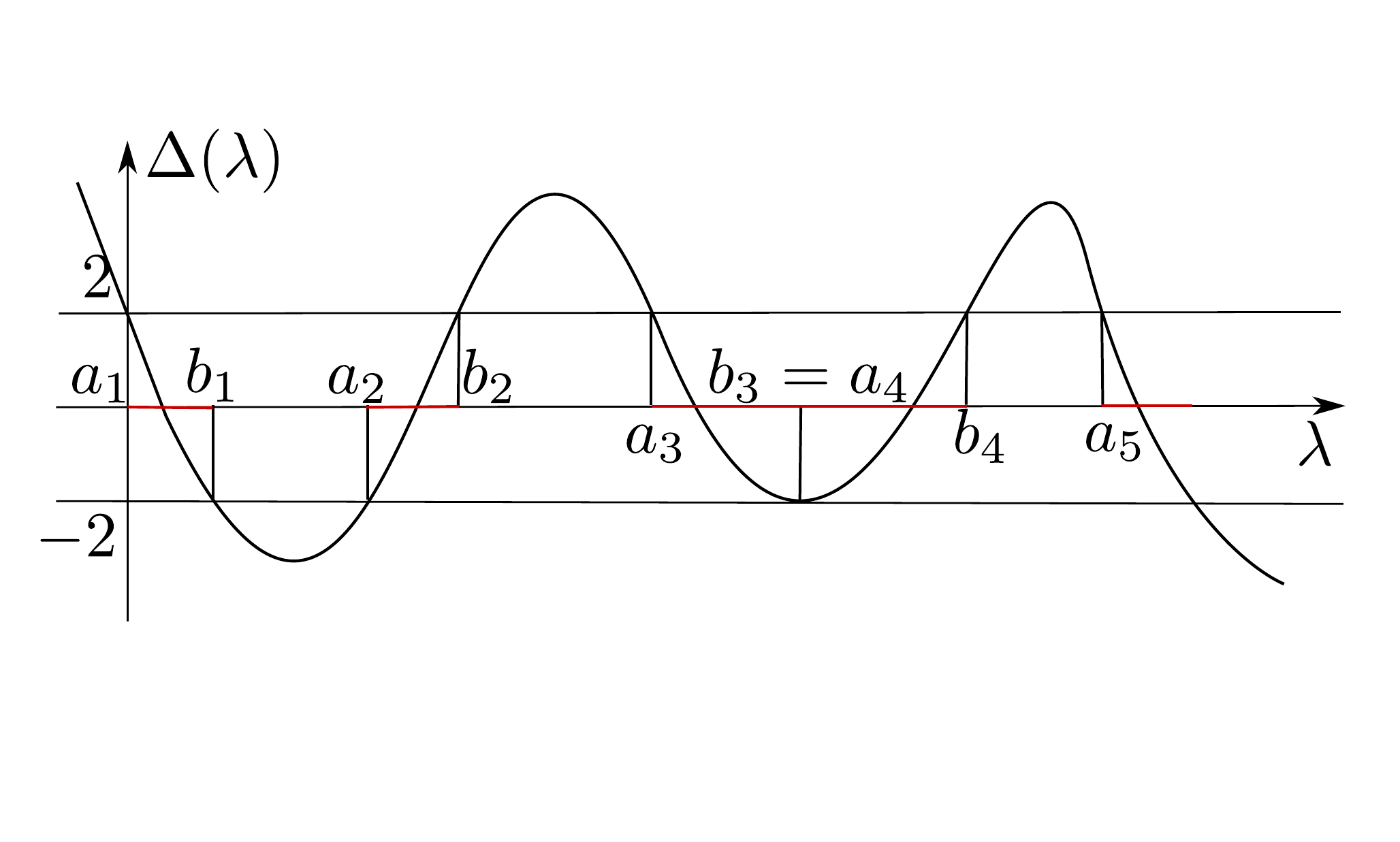}
\caption{Stability zones (spectral bands), shown in red, are separated by the instability zones (spectral gaps)}.
\end{center}
\eef
\ed

\br \label{R:zones_Lyap} Thus, we conclude that the spectral bands are characterized by the quasimomentum being real and Lyapunov exponent being equal to zero.
Correspondingly, the spectral gaps are characterized by the complex quasimomentum and positive Lyapunov exponent.
\er

\subsubsection{Spectrum of the Hill operator}
So far, the name ``spectral band'' was not explicitly related to the spectrum of the Hill operator $H=-d^2/dx^2 +V(x)$. It is not hard to show (which will be explained later on in the higher dimensions), that the spectrum $\sigma(H)$ indeed coincides with the union of all spectral bands. E.g., existence of bounded generalized solutions inside the spectral bands allows one to show that the bands do belong to the spectrum. Indeed, cutting off these solutions at infinity provides approximate eigenfunctions. On the other hand, one can show that inside the gaps there is no spectrum.

We now collect some important features of the spectrum:
\bt\label{T:1Dspect}\indent
\begin{enumerate}
\item There exists a sequence of real numbers
 \be\label{E:ajbj}
 a_1<b_1\leq b_2 <a_2 \leq a_3 < .... \mapsto \infty
 \ee
 (ends of the spectral bands) such that the spectrum $\sigma(H)$ coincides with the union of all spectral bands $I_j$
 \be\label{E:1dUnionbands}
 \sigma(H)=\bigcup\limits_{j\in\Z^+} I_j,
 \ee
 where
 \be
 I_1=[a_1,b_1], I_2=[b_2,a_2], I_3=[a_3,b_3], ....
 \ee
 \item The spectral bands have finite lengths and \textbf{do not overlap}, but might touch (see the proof in Lemma \ref{L:nooverlap}).
 \item \textbf{Generically (in the Baire category sense), w.r.t. to the $C^\infty$ potential, all gaps are present (open)}, i.e. there are no equalities in (\ref{E:ajbj}), so the bands do not touch \cite{Simon_gaps}.
\item \textbf{Finite gap (finite zone) potentials}, i.e. those that lead to finitely many open gaps only, are very special and can be all described \cite{Nov_sol84,Nov_sol94,DubKrich_book,DubrKrichNovVINITI}. However, they form a dense set in the space of all $C^\infty$-potentials \cite{MarchOstr,MarchOstr_corr,ColinKappelerDoubleHill}.
\item There are no open gaps if and only if the potential is constant (the famous \textbf{Borg's theorem} \cite{Borg}).
\item The rate of the \textbf{gaps' sizes $g_n$ decay} when $n\to\infty$ determines smoothness of the potential \cite{MarcOst,MityaginDzhakov,DM09,DM_jfa}. For instance, $C^\infty$ potentials correspond to the gap size decaying faster than any power of $1/n$. See, e.g. \cite[pp. 96--97]{DM09} for detailed history.
\item There are \textbf{isospectral} potentials. The sets of isospectral finite-zone potentials form tori \cite{Nov_sol94,mckean_inv75,McKean_Trub}.
\end{enumerate}
\et

\subsubsection{Dispersion relation}
We will introduce now the important notion of the dispersion relation for the Hill operator. It is rarely discussed in mathematics literature devoted to periodic ODEs, but it is central in solid state physics and crucial for the higher dimension considerations (see Firsova \cite{Firs75,Firs2000} on the Riemann surface of quasimomentum techniques in 1D).

The dispersion relation describes the spectral parameter $\lambda$ as a (multiple valued) function of the crystal momentum $k$:
\bd\label{D:disp1d}
The \textbf{dispersion relation} (or the \textbf{Bloch variety} $B_H$) of the Hill operator $H$ is the subset of $\R_k\times\R_\lambda$ defined as follows:
\be\label{E:1dDispBloch}
\ba{l}
B_H:=\{(k,\lambda)\in \R^2\, |\, Hu=\lambda u \mbox{ has a non-trivial Floquet-Bloch}\\
 \mbox{solution }u(x)=e^{ikx}p(x)\mbox{ with the quasi-momentum } k \}
\ea
\ee
The \textbf{complex dispersion relation} (or \textbf{complex Bloch variety}) $B_{H,\C}$ is defined analogously, only allowing both $k$ and $\lambda$ to be complex.
\ed
One immediately obtains the following statement:
\bt\label{L:dispr_per_even}\indent
\begin{enumerate}
\item The dispersion relation is $2\pi$-periodic with respect to $k$.
\item The real dispersion relation is even with respect to $k$. The complex one is invariant with respect to the mapping $(k,\lambda)\mapsto (-k,\overline{\lambda})$. (Indeed, due to the potential being real, the complex conjugate of a Floquet solution is also a Floquet solution.)
\item The spectrum of the operator $H$ coincides with the projection of its (real) dispersion relation onto the spectral $\lambda$-axis.
\item If the branches of the multiple valued function $k\in\R \mapsto \lambda\in\R$ are labeled in increasing order ($\lambda_1(k)\leq \lambda_2(k)\leq ...$), then the range of the $j$th branch is the $j$th spectral band of $H$.
\item In terms of the discriminant, the dispersion relation (both in the real and complex incarnations) is described as the set of all (correspondingly real or complex) solutions $(k,\lambda)$ of the equation
\be\label{E:dispeq1d}
\Delta(\lambda) - 2\cos k =0.
\ee
In other words, it is the graph of the multiple-valued function
\be \label{E:dispeq1dexpl}
\lambda=\Delta^{-1}(2\cos k).
\ee
\end{enumerate}
\et
Formula (\ref{E:dispeq1dexpl}) is less useful than (\ref{E:dispeq1d}), due to its multiple valued (and branching) nature. On the other hand, the function $F(\lambda,k):=\Delta(\lambda) - 2\cos k$ is an entire function in $\C^2$ of a finite exponential order (equal to $1$ in the Hill's case).
In particular, one can make the following observation, which gains prominence in PDE situation:
\bp\label{Analyt1d}
The dispersion relation of the Hill operator is a principal (i.e., of co-dimension one) analytic subset in $\C^2$.
\ep

Let us look at the case of the free operator. Then the dispersion relation boils down to $\cos\sqrt{\lambda}=\cos k$, or to the union of infinitely many parabolic branches
\be\label{Free_disp1d}
\lambda = (k+2n\pi)^2, n\in\Z,
\ee
see Fig. \ref{F:freeDisp_periodic}.
\bef[ht!]
\begin{center}
\includegraphics[scale=0.7]{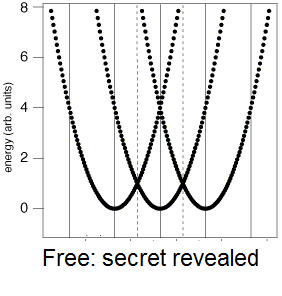}
\caption{The dispersion relation (over the whole quasi-momentum line) of the free operator.}\label{F:freeDisp_periodic}
\end{center}
\eef
Due to the periodicity with respect to $k$, it is customary to draw the dispersion relation only over the interval $[-\pi,\pi]$ (later on, this interval will acquire the name of a \textbf{Brillouin zone}). In fact, due to the evenness, one can, without loosing any information, restrict to $[0,\pi]$ (\textbf{reduced Brillouin zone}) only (see Fig. \ref{F:freeDisp}).
\bef[ht!]
\begin{center}
\includegraphics[scale=1.0]{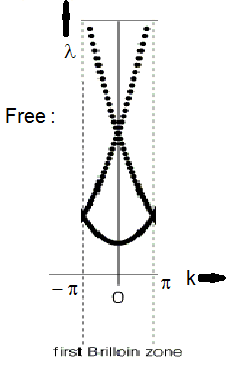}
\caption{The dispersion relation of the free operator.}\label{F:freeDisp}
\end{center}
\eef

One can see that the seeming complexity of the dispersion relation even in the free case, is an illusion, since one observes a single period view of a periodic sequence of parabolas.

Let us see how the picture gets perturbed when a small periodic potential $V(x)$ is turned on (Fig. \ref{F:nonfreeDisp}).
\bef[ht!]
\begin{center}
\includegraphics[scale=1.0]{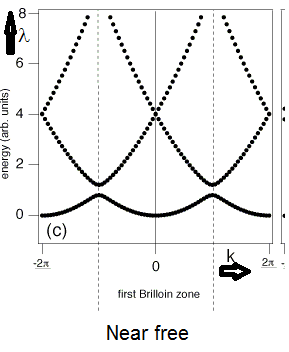}
\caption{The dispersion relation in presence of a potential.}\label{F:nonfreeDisp}
\end{center}
\eef
You see that under this perturbation some gaps are opening. However, it seems that over $[0,\pi]$ each branch is still monotonic. Let us see why this is indeed true:
\bl\label{L:monot}\indent
\begin{enumerate}
\item Each branch $\lambda_j(k)$ of the dispersion relation is monotonic over $[0,\pi]$.
\item In particular, the spectral band edges occur only at $k=0$ and $k=\pi$, i.e. for periodic or anti-periodic problems for the Hill operator.
\end{enumerate}
\el
\begin{proof} Indeed, if this were not true, taking into account the evenness with respect to $k$, one would have situation like in Fig. \ref{F:monotone}.
\bef[ht!]
\begin{center}
\includegraphics[scale=0.9]{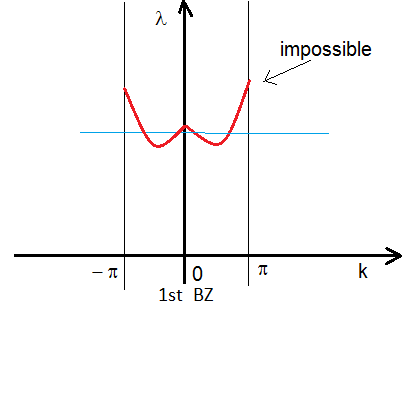}
\caption{Non-monotonicity is impossible.}\label{F:monotone}
\end{center}
\eef
If we pick the value of $\lambda$ at the level shown, we will find four values of $k$ where this level is reached. This means that the equation $Hu=\lambda u$ has four independent solutions. The ODE being of the second order, this is impossible.
\end{proof}

The same argument shows impossibility of band overlaps, as in Fig. \ref{F:impossible}:

\bef[ht!]
\begin{center}
\includegraphics[scale=0.6]{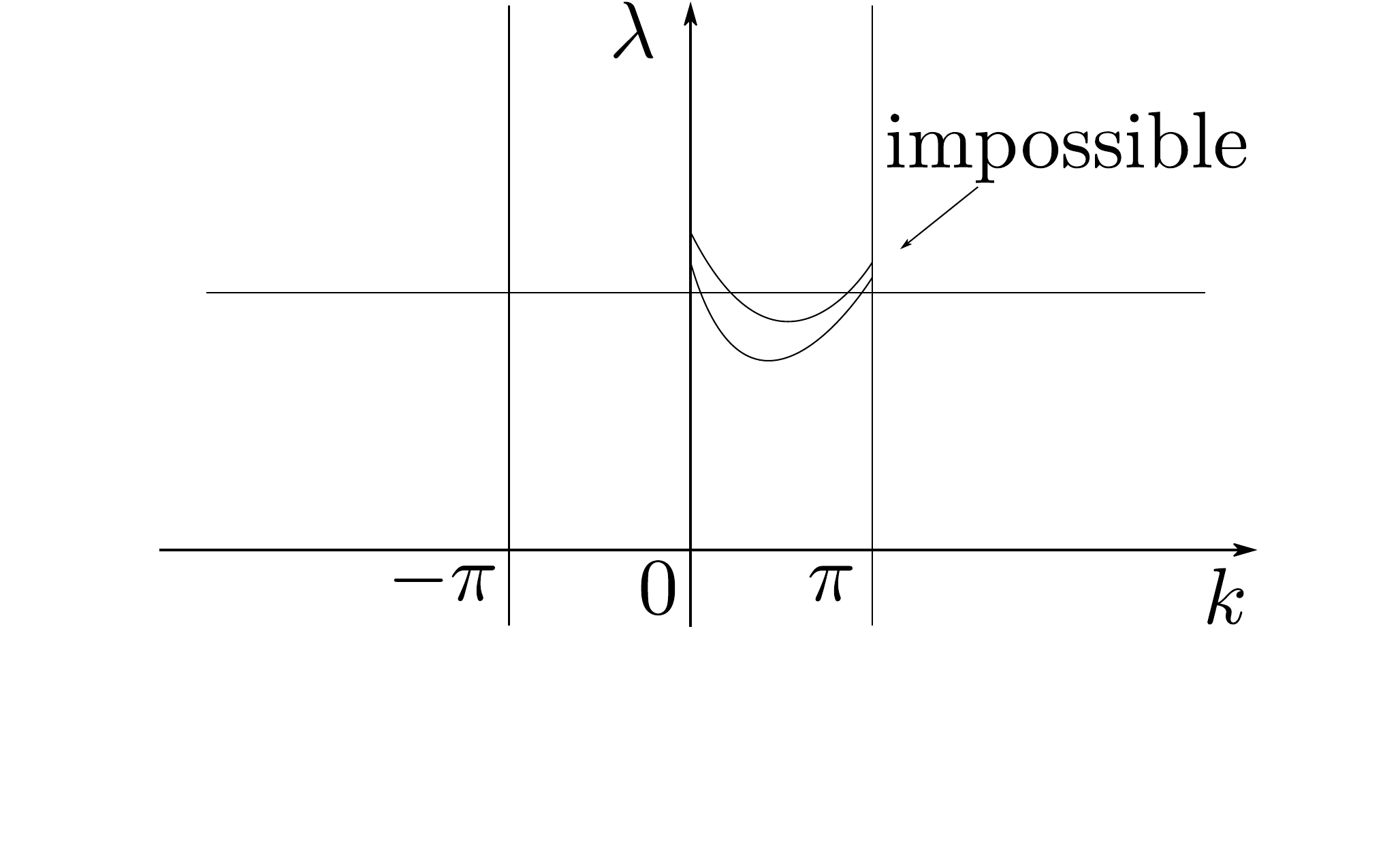}
\caption{No band overlap.}\label{F:impossible}
\end{center}
\eef

Thus, one concludes:
\bl\label{L:nooverlap}
The spectral bands of the Hill operator do not overlap (although they might touch).
\el

\br\label{R:only1d}
Notice that the conclusions about monotonicity, band edges occurring at $k=0$ and $k=\pi$ only, and absence of band overlap were drawn from the fact that the second order ODE cannot have more than two independent solutions. One can wonder, whether this is just an artifact of our proof, or something deeper. For instance, this counting cannot be used for PDEs. And indeed, we will see that for PDEs all these claims are in general incorrect.
\er

We formulate now without proofs some additional spectral properties:
\bt\label{T:extra1d}\indent
\begin{enumerate}
\item The pure point and singular continuous spectra of the Hill operator are empty. Thus, the spectrum is absolutely continuous\footnote{We will discus this at length for the PDE case.} \cite{Eastham_periodic,ReedSimon_v14,Kuc_floquet}.
\item The dispersion relation, as an analytic set, is irreducible (modulo the $2\pi$ shifts of the quasi-momentum). In other words, any open part of the dispersion relation uniquely determines the whole of it. To put it still differently, the smooth part of $B_H$ is connected \cite{Kohn} (see also \cite{AvrSim_ap78}). (See \cite{Papanic_ODEperiodic} for additional irreducibility results for periodic ODEs.)
\item The Bloch variety (which is analytic, as we know), is generically not algebraic (e.g., \cite{mckean_inv75,McKean_Trub}).
\end{enumerate}
\et

\section{Lattices in $\R^n$}\label{S:lattices}
Switching to the multi-periodic case, we need first to go through some basics concerning lattices. One can find a nice more detailed introduction in \cite{Skr_psim85}.

\bd\label{D:lattice}
A (Bravais) \textbf{lattice} $\G$ in $\R^n$ is the set of all integer linear combinations of $n$ linearly independent vectors $a_1, ... ,a_n$ (see Fig. \ref{F:lattice}:
\be\label{E:lattice}
\G=\{\g\in\R^n\, |\, \g=\sum\limits_{j=1}^n \g_ja_j, \g_j\in\Z\}.
\ee
\ed
We will identify $\G$ with the corresponding group of shifts of $\R^n$.
\bef[ht!]\begin{center}
\includegraphics[scale=0.4]{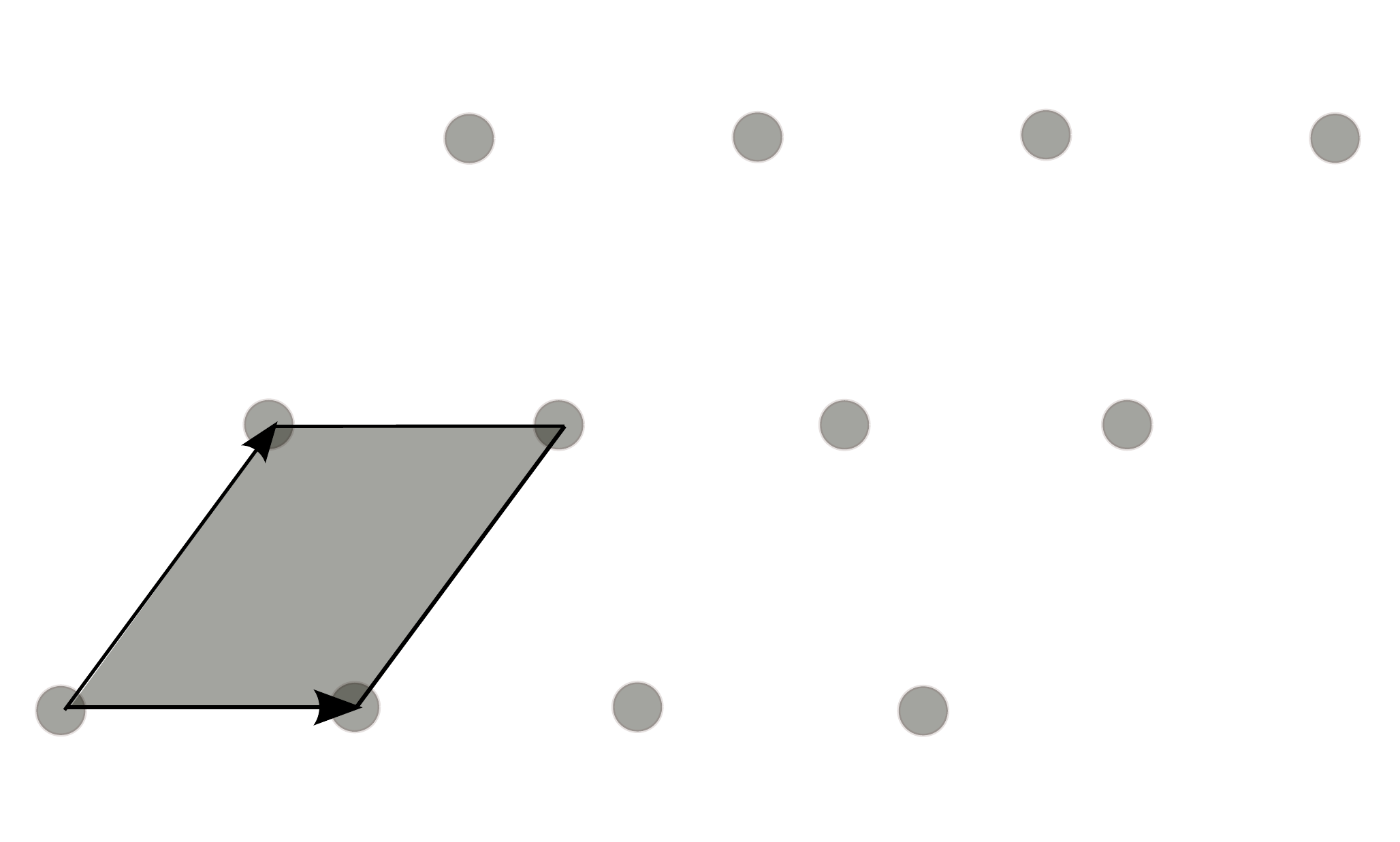}
\caption{A lattice and its fundamental domain (shaded).}\label{F:lattice}
\end{center}
\eef

\bd\label{D:duallat}
The \textbf{dual (reciprocal) lattice (see Fig. \ref{F:dlattice}) to $\G$}, $\G^* \subset(\R^n)^*$ is defined as follows:
\be\label{E:duallat}
\G^*=\{k\in(\R^n)^*\, |\, \langle k,\g\rangle\in 2\pi\Z \mbox{ for any }\g\in\G\}.
\ee
The original lattice $\G$ is sometimes called the \textbf{real lattice}.
\ed
In particular, when $\G=\Z^n$ and the duality is coming from the standard Euclidean scalar product, then $\G^*=2\pi\Z^n$.
\bef[ht!]\begin{center}
\includegraphics[scale=0.5]{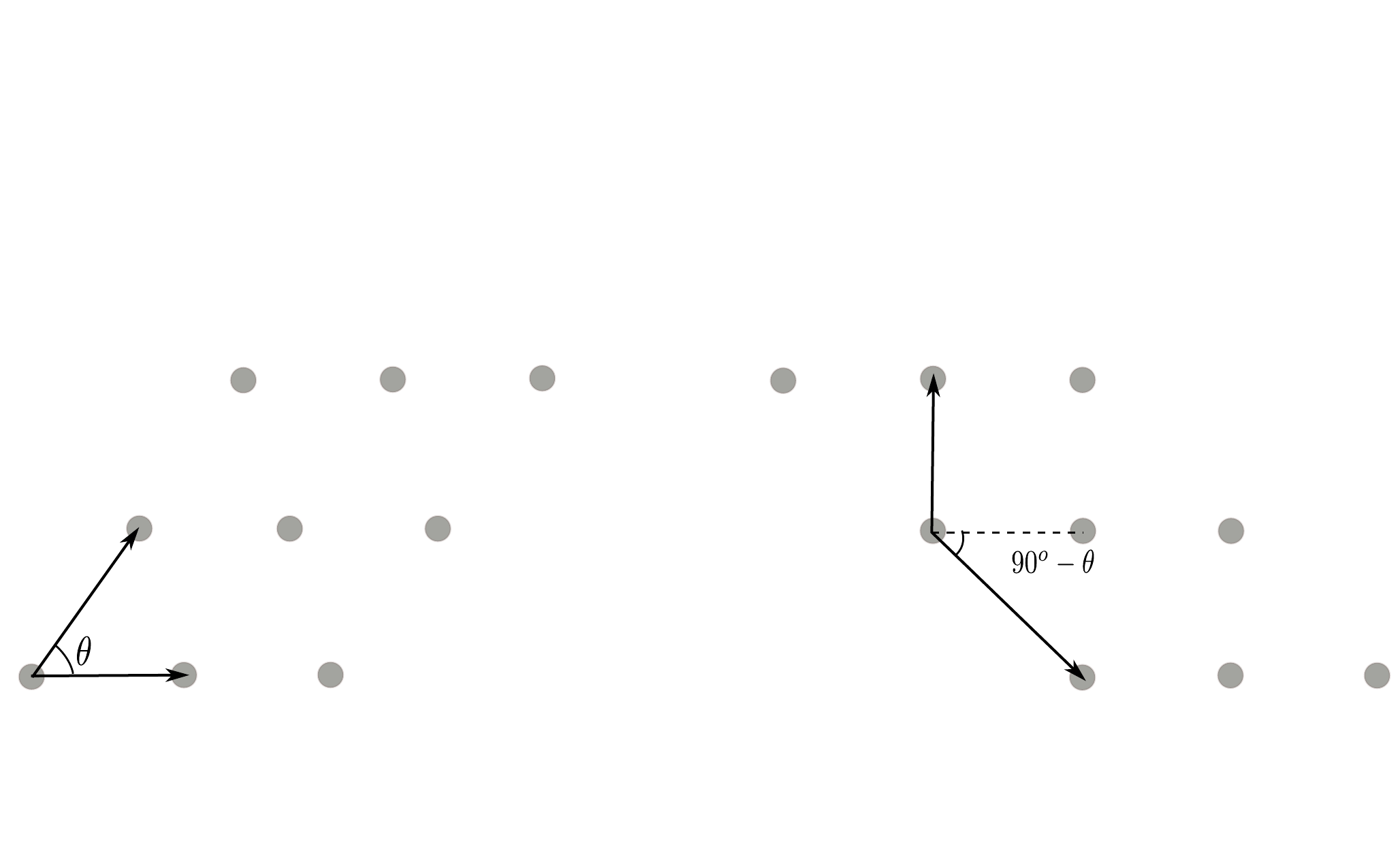}
\caption{A real and its reciprocal lattices.}\label{F:dlattice}
\end{center}
\eef

\bd\label{D:WS_Br}
We fix a connected \textbf{fundamental domain} $W$ (\textbf{Wigner-Seitz cell}) of the (real) lattice $\G$ in $\R^n$, i.e. a closed domain such that its $\G$-shifts may intersect only along their boundaries and cover the whole space (See Fig. \ref{F:lattice})\footnote{A fundamental domain is clearly not defined uniquely.}.

We will also fix a fundamental domain $\B$ of (the reciprocal lattice) $\G^*$ in $(\R^n)^*$, which we will call the \textbf{Brillouin zone}. One of the options (standard in physics) is to choose the \textbf{Voronoi cell}, i.e. the set of all points such that the origin is the closest point to them in $\G^*$ (see Fig. \ref{F:brill}). In this work, we'll be choosing a parallelepiped as the fundamental domain (see below) and also skip the consideration of important so called second, third, etc. Brillouin zones (see, e.g. \cite{Skr_psim85,Col_msmf91} for definitions and interesting discussions).
\ed
\bef[ht!]\begin{center}
\includegraphics[scale=0.5]{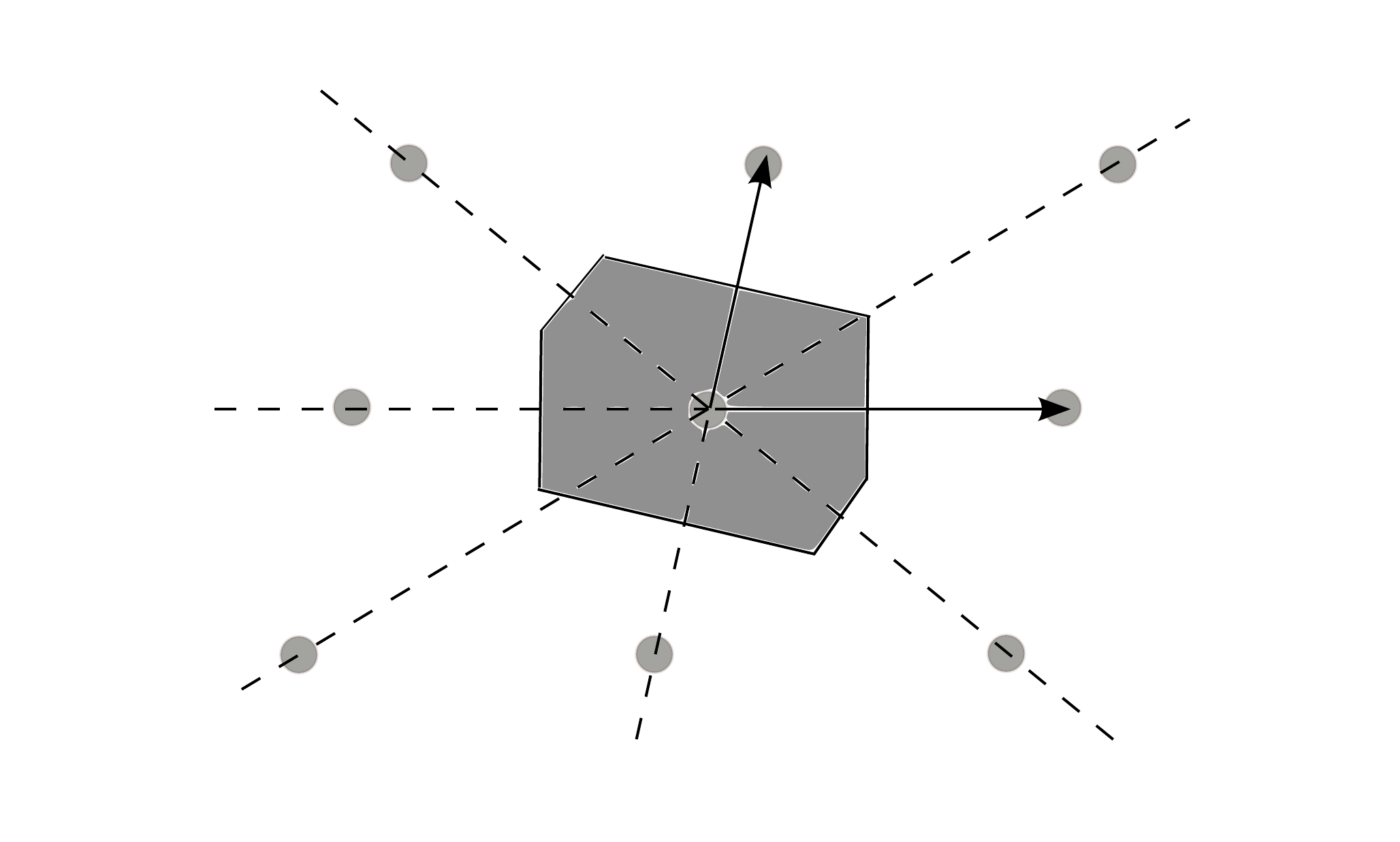}\\
\end{center}\caption{A Brillouin zone of a lattice.}\label{F:brill}
\eef

The particular choice of a lattice does not matter in many, and does matter in some results, but for the sake of simplicity, we will assume, unless noted otherwise, that
\be\label{E:standardchoice}
\G=\Z^n, \G^*=2\pi\Z^n, W=[0,1]^n, \B=[-\pi,\pi]^n.
\ee

We also introduce two \textbf{tori} that correspond to the two lattices
\be\label{E:tori}
\T:=\R^n/\G \mbox{ and }   \T^*:=(\R^n)^*/\G^*,
\ee
equipped with normalized Haar measures on both. E.g., under our standard choice (\ref{E:standardchoice}), the measures are correspondingly $dx$ and $(2\pi)^{-n}dk$.

Note that under our assumption that $\G=\Z^n$, the torus $\T^*$ can be considered as the unit torus in $\C^n$:
\be\label{E:unittorus}
\T^*=\{(z_1,...,z_n)\in\C^n\,|\, |z_1|=...=|z_n|=1\}.
\ee

As usual, $\G$ - ($\G^*$ -) periodic functions on $\R^n$ (on $(\R^n)^*$) can be identified with  functions on the torus $\T$ ($\T^*$).

\textbf{Fourier series (FS)} identify $L^2(\T)$ with the $l^2$-space on $\G^*$:
\be\label{E:FS}
f(x) \mapsto \{f_k:=\int_\T f(x)e^{-ik\cdot x}dx\}_{k\in\G^*} .
\ee
Analogously with  $L^2(\T^*)$ and $l^2$ on $\G$.

One can also consider vector-valued Fourier series. Namely, let $\H$ be a Hilbert space and $L^2(\T,\H), \, l^2(\G^*,\H)$ be the spaces of $\H$-valued $L^2$- functions on $\T$ and $\G^*$ correspondingly and the Fourier Series expansion is defined as in (\ref{E:FS}). The following theorem restates standard results concerning Fourier series:
\bt\label{T:FSPlanchPW}\indent
\begin{itemize}
\item The FS expansion is an isometry of $L^2(\T,\H)$ onto $l^2(\G^*,\H)$.
\item A function $\in L^2(\T,\H)$ is infinitely differentiable if and only if the norms $\|f_k\|$ of its Fourier coefficients decay faster than any power of $k$.
\item A function $\in L^2(\T,\H)$ allows analytic continuation into a complex neighborhood of $\T$ (see (\ref{E:unittorus})) if and only if the norms $\|f_k\|$ of its Fourier coefficients decay exponentially.
\end{itemize}
\et
\section{Periodic operators}

From now on, we will be interested in studying linear elliptic partial different operators with periodic coefficients. Our main ``test'' example is the Schr\"odinger operator in $L^2(\R^n)$:
\be\label{E:sch}
H=-\Delta + V(x),
\ee
with a ``sufficiently nice'' real electric potential $V$, periodic with respect to the group $\G=\Z^n$  (and thus the dual lattice is $\G^*=2\pi\Z^n$). The domain of $H$ is the Sobolev space $H^2(\R^n)$. In most cases, the reader will not be misled thinking just of this operator.

Many other periodic operators of mathematical physics arise in applications and need to be studied. The techniques described in this survey work (sometimes with some caveats) for them as well.  We will from time to time address some of those, but so far we just briefly describe some examples worthy of studying (and being studied):

\begin{itemize}
\item Although we assumed above a ``nice'' periodic potential $V(x)$ (continuous or even smooth), this condition is unnecessarily strong (e.g., $L_{2,loc}$ usually suffices). However, the main issues already arise in the smooth case. So, unless specified otherwise, \textbf{coefficients in all operators will be assumed sufficiently smooth (e.g., $C^\infty$)}, albeit in practical applications they often are non-smooth, even discontinuous.
\item The Schr\"{o}dinger operator (\ref{E:sch}) is self-adjoint, however \textbf{many of the techniques and results do not require self-adjointness.} For instance, when we discuss in Section \ref{SS:ac,pp,sing} absolute continuity of the spectrum, we will restate it as absence of the pure point spectrum. In this form, it holds, without any change in the proof, in the non-self-adjoint setting as well.
\item More general Schr\"odinger operators are also of interest and present more difficulties in studying them. These are, first of all,
    the \textbf{magnetic Schr\"odinger operator}
    $$(\frac{1}{i} \nabla - A(x))^2 +V(x)$$
    with periodic magnetic and electric potentials $A(x)$ and $V(x)$ (one assumes the gauge $\nabla\cdot A=0$).

    Even more difficulties one encounters in \textbf{presence of periodic metric:}
    $$-\nabla \cdot g(x)\nabla +iA(x)\cdot\nabla +V(x),$$
    where $g(x)$ is a periodic positive definite matrix-function.
\item Periodic elliptic \textbf{operators of higher than second order} are also of some interest. Here, one should beware that such operators, unlike the 2nd order ones, might not obey even the weakest uniqueness of continuation laws, which influences the validity of some of the results (e.g., absence of the pure point spectrum).
\item Periodic elliptic \textbf{systems} (including overdetermined ones) can also been considered, first of all the \textbf{Maxwell operator} (in its 2nd order incarnation):
    \be\label{E:maxwell}
    \nabla \times \varepsilon^{-1}(x) \nabla \times,
    \ee
where $\varepsilon (x)$ is a periodic positively definite scalar function (or tensor) and $\nabla \times$ is the $curl$ operator.
Here one encounters an additional difficulty, since this operator is not elliptic by itself, but only as a \textbf{member of an elliptic complex} of operators.

\item The natural mapping $\R^n \mapsto \T$ is a normal \textbf{abelian covering} of the torus $\T$.
The techniques that we will consider, as well as many results, apply to \textbf{periodic operators on coverings}
$$M \mapsto N,$$
subject to the following three conditions: (1) The base $N$ is compact. (2) The deck group $\G$ of the covering is finitely generated and virtually abelian\footnote{I.e., contains a finite index abelian subgroup.}. (3) The operator on $M$ is ``elliptic''   in the sense that being pushed down to $N$, it is a Fredholm operator in appropriate spaces.

In particular, \textbf{periodic operators on abelian coverings of compact} \textbf{Riemannian manifolds}, \textbf{analytic manifolds}, and even \textbf{graphs}, succumb gladly to the theory.
\item \textbf{Elliptic periodic boundary value problems} arise in various applications (e.g., waveguides and photonic crystals). Here periodicity is imposed on the shape of the domain, coefficients of the operator, and boundary conditions. Some issues easily resolved for operators in the whole space become much harder in this new setting (see, e.g. Section \ref{SS:guides}).
\item Ellipticity condition can be weakened to \textbf{hypoellipticity, e.g. parabolicity}, although the known results are much weaker here (see Section \ref{SS:evolution} and \cite[Ch. 5, and references therein]{Kuc_floquet}).
\item \textbf{Non-hypoelliptic periodic equation}s, e.g. important time periodic \textbf{hyperbolic or non-stationary Schr\"odinger} equations \textbf{require different techniques}, due to lack of Fredholm property (see, e.g. \cite{YajimaLargeTime,YajimaScat,YajimaSurv} and Section \ref{SS:evolution}).
\item Some, but not all, studies carry over to periodic elliptic \textbf{pseudo-differential operators} ($\Psi$DOs) (see, e.g., \cite[Section 3.4.C]{Kuc_floquet}, \cite{Sobolev_INI,ParnSob_Invent}, and references therein).

\end{itemize}

\section{Floquet transform and direct integral decomposition}\label{S:Fltrans}

Our main tools in 1D were the spectral analysis of the monodromy operator and the Lyapunov reduction theorem. Both of them rely upon the propagator along the solutions of the Cauchy problem. This raises some hopes that an analog of the 1D Floquet theory \textbf{might} work for time-periodic PDEs of evolution type (parabolic, hyperbolic, non-stationary Schr\"odinger), where such propagators are nicely defined. And indeed, {\bf sometimes} this does happen, although, surprisingly (and shamefully) some basic issues remain still unresolved (see some details in Section \ref{SS:evolution}). However, for non-evolution (e.g., elliptic) periodic PDEs using such propagators is all but impossible \includegraphics[scale=0.08]{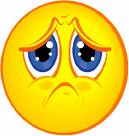}. One thus has to resort to a different technique, which we outline below.

\subsection{Floquet transform}
As we have recalled above,
Fourier series transform functions of $\g\in\G$ (even with values in a Hilbert space) into functions of $z\in\T^*$:
$$
\{f_\g\}_{\g\in\G}\mapsto \widehat{f} (z):=\sum_{\g\in\G} f_\g z^\g= \sum_{\g\in\G} f_\g e^{ik\cdot\g},
$$
with standard $L^2$ isometry and Paley-Wiener type theorems being preserved.
Here $k\in\R^n$ and $z$ belongs to the unit torus $\T^*$.

Let now $f(x)$ be a function in $L^2_{loc}(\R^n)$.   We cover $\R^n$ by the shifted copies of the fundamental domain $W$:
\be
\R^n=\bigcup\limits_{\g\in\G} (W-\g),
\ee
cut $f$ into the corresponding pieces $f|_{W-\g}$
and then shift them all back to the original $W$:
\be\label{E:pieces}
f_\g (x) := f|_{W-\g} (x-\g).
\ee
If we restrict here $x$ to being in the Wigner-Seitz cell $W$, we get a function on $\G$ with values in the Hilbert space $L^2(W)$:
\be\label{E:vectorvalued}
\g\in\G \mapsto f_\g \in L^2(W).
\ee
Now summing the Fourier series, we arrive to our main tool:
\bd\label{D:FT}
The \textbf{Floquet transform}\footnote{Neither the name ``Floquet transform'' is commonly accepted, nor Floquet introduced this transform. Various other names have been suggested: Bloch transform, Gelfand transform (due to \cite{Gel_dansssr50}), Zak transform, and probably some others. Here we see an instance of the so called \textbf{Arnold's principle}:  \emph{If a notion bears a personal name, then this name is not the name of the discoverer} \cite{ArnTeach}. (See also
the \textbf{Berry's Principle}: \emph{the Arnold's Principle is applicable to itself} and \textbf{Stigler's law of eponymy} \cite{Stigler}.) However, it is handy to use a name, rather than point a finger to a formula. Thus, the author chose Floquet, since this transform is very natural for Floquet theory.} $\U$ acts as follows:
\be\label{E:FT}
f(x)\mapsto \U f(x,k):=\sum_{\g\in\G} f_\g(x) e^{ik\cdot\g}= \sum_{\g\in\G} f_\g(x)z^\g.
\ee
I.e., this is just the sum of the Fourier series, whose coefficients are the pieces of $f$ over the shifted copies of the fundamental domain.
\ed
Here, as before, $z=(z_1,...,z_n):=(e^{ik_1}, ... ,e^{ik_n})$ is the Floquet multiplier corresponding to the crystal momentum $k=(k_1,...,k_n)$. We will also abuse notations, writing $$\U f(x,z)=\sum_{\g\in\G} f_\g(x)z^\g$$.

It will be convenient to extend $\U f(x,k)$ from $x\in W$ to the whole $\R^n$ just by removing the restriction to $W-\g$ in (\ref{E:pieces}), which results in
\be\label{E:FTall}
f(x)\mapsto \U f(x,k):=\sum_{\g\in\G} f(x-\g) e^{ik\cdot\g}= \sum_{\g\in\G} f(x-\g)z^\g.
\ee
If this is done, then one observes the following useful properties:
\begin{eqnarray}
\U f(x+\g,k)=\U f(x,k)e^{ik\cdot \g}, \g\in\G ,\label{E:autom}\\
\U f(x+\g,z)=\U f(x,k)z^{\g}, \g\in\G ,\label{E:autom2}\\
\U f(x,k+k')=\U f(x,k), k'\in\G^*.\label{E:k-periodic}
\end{eqnarray}
The first two equalities, (\ref{E:autom}) and (\ref{E:autom2}), say that, being considered on the whole space $\R^n$, $\U f(x,k)$ is \textbf{$\G$-automorphic} with respect to $x$ with the character $e^{ik\cdot \g}$ (this is also called \textbf{cyclic property, quasi-periodicity, or Floquet property}). The last one, (\ref{E:k-periodic}) claims \textbf{$\G^*$-periodicity} with respect to $k$.

It is easy to get an inversion formula for the Floquet transform. Indeed, it is enough to find the Fourier coefficients of $\U f(x,k)$ with respect to $k$ and place them where they belong, i.e. on the shifted copies of the fundamental domain. This leads to the following inversion formula:
\be\label{E:invFT1}
f(x)=\int_{\T^*}\U f(x,k)e^{-ik\cdot \g}dk,\, x\in W-\g,\mbox{ for any }\g\in\G.
\ee
On the other hand, if we consider the Fourier coefficients $f_\g$ as the shifted copies of $f$ in the whole space $\R^n$, then just the zero's coefficient recovers the whole function:
\be\label{E:invFT2}
f(x)=\int_{\T^*}\U f(x,k)dk,\, x\in \R^n.
\ee

\subsection{Plancherel and Paley-Wiener type theorems}
Now, application of Theorem \ref{T:FSPlanchPW} to $\H=L^2(W)$ leads to the following range theorems for the Floquet transform:
\bt\label{E:FTL2planchPW}\indent
\begin{itemize}
\item $\U$ is isometry from $L^2(\R^n)$ onto $L^2(\T^*, L^2(W))$.
\item $\U f(\cdot,k)$ is infinitely differentiable on $\T^*$ as a function with values in $L^2(W)$ iff the norms $\|f\|_{L^2(W+\g)}$ decay when $|\g|\mapsto \infty$ faster than any power of  $|\g|$.
\item $\U f(\cdot,k)$ is analytic in a complex neighborhood of $\T^*$ as a function with values in $L^2(W)$, iff the norms $\|f\|_{L^2(W+\g)}$ decay exponentially fast when $|\g|\mapsto \infty$.
\end{itemize}
\et
Dealing with differential operators, we will be also interested in the behavior of Sobolev spaces $H^s (\R^n)$ under the Floquet transform. This seems to be a piece of cake, just replacing everywhere $L^2$ with $H^s$. Well, not so fast:
\bp\label{P:rangeSobolev}
$\U$ is an isometry from $H^s(\R^n)$ \textbf{onto a proper subspace} of $L^2(\T^*, H^s(W))$.
\ep
Oops! What is wrong with Sobolev spaces? The answer is not hard to figure out. Indeed, if one takes an arbitrary function $F(x,k)$ in $L^2(\T^*, H^s(W))$, its alleged pre-image under the Floquet transform can be, as we have figured out before, recovered by taking Fourier coefficients of $F$ and distributing them to the appropriate shifts of the fundamental domain $W$. The resulting function is, by construction, in $H^s$ inside of each of the shifted copies of $W$. Moreover, the sum of squares of their $H^s$ norms is finite. The problem is that there is nothing enforcing the smoothness across the boundary between two adjacent cells.

Fortunately, it is not hard to fix this and describe the range of the transform explicitly. Clearly some boundary conditions are needed to glue the pieces together. And indeed these are easy to establish for instance when $s=2$ (or any $s\in\Z^+$). It is, however, easier to avoid this and to adopt a much more telling (and applicable for any $s\geq 0$) view. The clue on how to do this comes from the automorphicity property (\ref{E:autom}, \ref{E:autom2}).
We introduce the following
\bd\label{D:automSobolev}
For any $k\in\C^n$, the space $H^s_k(W)$ is the space of restrictions to $W$ of $H^s_{loc}$-function on $\R^n$ that are automorphic with the given $k$:
\be\label{E:Hskdef}
H^s_k(W):=\{f|_{W}\, | \,f\in H^s_{loc} (\R^n), f(x+\g)=e^{ik\cdot \g}f(x) \mbox{ a.e., } \forall  \g\in\G\}.
\ee
\ed
Another useful object is a $k$-dependent smooth line bundle over $\T$. Let us fix $k\in\C^n$ and consider first the trivial line bundle $E:=\R^n \times \C$ over $\R^n$. We define a ($k$-dependent) $\G$-action on $E$ as follows:
\be\label{E:Ebundle}
\mbox{For any }\g\in\G \mbox{ and } (x,c)\in E, \tau_k(\g) (x,c):=(x+\g, e^{ik\cdot x} c).
\ee
\bd\label{D:Ek}
We denote by \textbf{$E_k$ the line bundle over the torus $\T$} obtained by factoring out the action $\tau_k$ of $\G$ on the bundle $E$.
\ed
Now the following claims are immediate to obtain:
\bp\label{P:Hskproperties}Let $s\geq 0$. Then, for any $k$,
\begin{enumerate}
\item $H^s_k$ is a closed subspace in $H^s(W)$.
\item $H^0_k=L^2_{k}=L^2(W)$.
\item $H^s_{k+k^\prime}=H^s_k$ for any $k^\prime \in \G^*$.
\item A function $u$ belongs to $H^s_k$ iff
    \be\label{E:A-k}
    u(x)=e^{ik\cdot x}v(x)\mbox{ with }v(x)\mbox{ being }\G-\mbox{periodic, i.e. } v\in H^s(\T).
    \ee
\item Moreover, $\H^s:=\bigcup_k H^s(k)$ is an analytic Banach vector sub-bundle of the trivial bundle $\C^n\times H^s(W)$ over $\C^n$.
\item There is a natural correspondence between $\G$-automorphic functions on $\R^n$ with a quasi-momentum $k$ and sections of the linear bundle $E_k$ over $\T$.
\item Under this correspondence, elements of $H^s_k$ correspond to $H^s$-sections of the bundle $E_k$.
\item The operator $H$ preserves the $\G$-automorphicity. In particular, $H$ defines an elliptic operator $H(k)$ on sections of the bundle $E_k$.
\end{enumerate}
\ep

\br\label{E:twistLapl}\indent
\begin{itemize}
\item In the particular case when $H$ is the Laplacian $-\Delta$, the operators $-\Delta(k)$ are called \textbf{twisted Laplacians}. We call $H(k)$ \textbf{twisted Schr\"odinger operators}, or (more common) \textbf{Bloch Hamiltonians}.
\item The advantage of dealing with the family of operators $H(k)$ rather than the single operator $H$ is that each $H(k)$, being an elliptic operator in sections of a bundle over a \textbf{compact} manifold $\T$, has purely discrete spectrum, unlike the operator $H$, whose spectrum is continuous.
\item For the operator $H=-\Delta +V(x)$, the corresponding $H(k)$ acts by the same differential expression on $W$ with the domain $H^2_k$, i.e. with the cyclic boundary conditions. In this view, the domain of the operator analytically rotates with $k$, while the differential expression of $H$ stays the same. One of its advantages is that the domain $H^s_k$ of the operator is $\G^*$-periodic with respect to $k$, so it can be considered as depending on the Floquet multiplier $z=(e^{ik_1}, ... ,e^{ik_n})$ only.
\item Alternatively, according to (\ref{E:A-k}), the operator $H(k)$ can be considered as acting, for all $k$, on periodic functions on $\T$. This is achieved by commuting with an exponential function:
    \be\label{E:commExp}
    e^{-ik\cdot x}\circ \left(-\Delta +V(x)\right)\circ e^{ik\cdot x}.
    \ee
    This leads to the following differential expression for $H(k)$
    \be\label{E:repr_polyn}
    -\Delta - 2ik\cdot \nabla +k^2+ V(x)
     \ee
     acting on the torus $\T$, i.e., on $\G$-periodic functions.
     The advantage of this representation is the explicit polynomial dependence of the differential expression on the quasi-momentum and the domain being fixed. On the other hand, explicit periodicity in $k$ is lost.

There are advantages and disadvantages of both representations. It is thus recommended to choose them judiciously, depending on the problem.
\item The latter representation can be generalized to any periodic differential operator $L(x,D)$, where $D:=\dfrac1i \dfrac{\partial}{\partial x}$:
    \be\label{E:repr_polyn_gen}
    L(k)=L(x,D+k).
     \ee
\end{itemize}
\er
We can now formulate the exact range theorem for the Floquet transform in Sobolev spaces:
\bt\label{T:RangeHs}
\indent
\begin{itemize}
\item $\U$ is isometry from $H^s(\R^n)$ onto the space $L^2(\T^*, \H^s)$ of the $L^2$-sections of the bundle $\H^s$.
\item The section $\U f(\cdot,k)$ is infinitely differentiable on $\T^*$ iff the norms $\|f\|_{H^s(W+\g)}$ decay when $\|\g\|\mapsto \infty$ faster than any power of  $\|\g\|$.
\item The section $\U f(\cdot,k)$ is analytic in a complex neighborhood of $\T^*$ iff the norms $\|f\|_{H^s(W+\g)}$ decay exponentially fast when $\|\g\|\mapsto \infty$.
\end{itemize}
\et

\subsection{Direct integral decomposition}\label{SS:directintegral}
We can summarize the results of previous sections as the following
\textbf{direct integral decompositions} of functional spaces and of the operator $H$:
\be\label{E:directInt}
L^2(\R^n)=\int^\bigoplus_{\T^*} L^2(W),\quad H^s(\R^n)=\int^\bigoplus_{\T^*} H^s_k, \quad H=\int^\bigoplus_{\T^*} H(k).
\ee
We will not get into any deeper discussions of the direct integral technique (see, e.g. \cite{ReedSimon_v14,Scarpellini_stability,FollandHarm,NielsenDirInt,CorwinGreenleaf,Dixmier}).

\section{Dispersion relation and all that}\label{S:DR}
As there is in general no analog of the 1D discriminant in higher dimensions, the dispersion relation takes the lead. Its definition and many properties are analogous to the ones in 1D, although some important distinctions do arise.

\subsection{Dispersion relation = Bloch variety}
\bd\label{D:disp1d}
The (real) \textbf{dispersion relation} (or the (real) \textbf{Bloch variety} $B_H$) of the periodic Schr\"odinger operator\footnote{The definition stays the same for other periodic operators,} $H$ is the subset of $\R^n_k\times\R_\lambda$ defined as follows:
\be\label{E:ndDispBloch}
\ba{l}
B_H:=\{(k,\lambda)\in \R^{n+1}\, |\, Hu=\lambda u \mbox{ has a non-trivial Bloch-Floquet }\\
 \mbox{solution }u(x)=e^{ik\cdot x}p(x)\mbox{ with the quasi-momentum } k \}
\ea
\ee
The \textbf{complex dispersion relation} (or \textbf{complex Bloch variety}) is defined analogously, only allowing both $k$ and $\lambda$ to be complex, i.e. $B_{H,\C}\subset \C^{n+1}$. As we will see later on, the ability of considering the \emph{complex} Bloch variety turns out to be very important.
\ed

\subsection{Dispersion relation and the spectrum}\label{SS:DRandSpect}
Let us assume that $k\in\R^n$. The operator $H(k)$ being a self-adjoint elliptic operator in sections of a linear bundle on the torus, implies

\bl\label{L:discretesp}
 For any $k\in\R^n$, the operator $H(k)$ is bounded below and has a discrete (real) spectrum
\be\label{E:bands}
\sigma(H(k)) =\{ \lambda_j(k)| \lambda_1(k)\leq \lambda_2(k)\leq ... \leq \lambda_d(k)\leq \dots \mapsto \infty\}.
\ee
\el

\bd\label{D:bandfunct}
The function $\lambda_j(k)$ is called the \textbf{$j$th band function}.
\ed
\br
Notice that for complex quasimomenta $k$, the spectrum of $H=-\Delta(k)+V$ is also discrete \cite[pp.180-190]{Agmon}, although labeling the band functions becomes an issue.
\er
The following statements held, due to the direct integral expansion (\ref{E:directInt}) of the operator, perturbation theory \cite{Kato_perturbation}, and standard properties of analytic Fredholm operator-functions (e.g., \cite{ZaiKreKucPan_umn75,Kuc_floquet,Wil_jam78}):
\bt\label{T:elempropdisp}\indent
\begin{enumerate}
\item The band functions are continuous and piece-wise analytic.
\item The graph of the multiple-valued mapping
 \be
 k\in\R^n \mapsto \sigma(H(k))
 \ee
coincides with the dispersion relation (Bloch variety) $B_H$ of $H$.
 \item The latter claim also holds in the complex case, i.e.
the graph of the multiple-valued mapping
 \be
 k\in\C^n \mapsto \sigma(H(k))
 \ee
coincides with the complex dispersion relation (complex Bloch variety) of $H$.
\item The dispersion relation is $\G^*$-periodic with respect to $k$ and thus it is sufficient to consider it only over the Brillouin zone $\B$.
\item The dispersion relation is symmetric (even) with respect to the mapping $k \mapsto -k$. (This fails if the potential is not real.)
\item The spectrum $\sigma(H)$ is the range of the (real) dispersion relation, i.e.
    \begin{eqnarray}
    \sigma(H)=\bigcup\limits_{k\in\B}\sigma(H(k))\label{E:ndspect}\\
    =\{\lambda\in\R\,|\, \exists k\in\R^n, \mbox{ such that }\lambda\in\sigma(H(k)) \}\label{E:ndspect1}\\
    =\{\lambda\in\R\,|\, \exists k\in\R^n, j\in\Z^+, \mbox{ such that }\lambda=\lambda_j(k) \}.\label{E:ndspect2}
    \end{eqnarray}
\end{enumerate}
\et
The last statement of the theorem is a very general claim about spectra of direct integrals of operators (see, e.g., \cite{ReedSimon_v14,Kuc_floquet}).

In Fig. \ref{F:ndNonfreeDisp} one finds an example of the dispersion relation for a non-zero potential:
\bef[ht!]
\begin{center}
\includegraphics[scale=0.35]{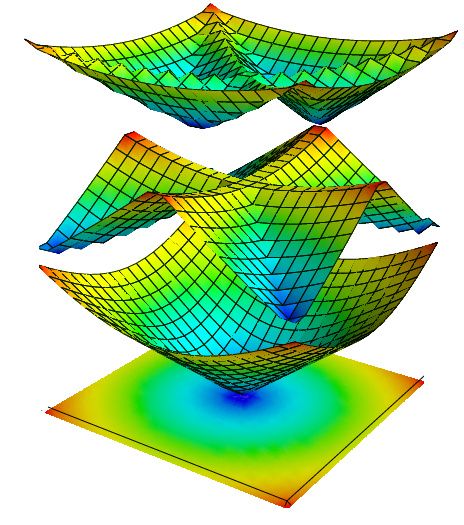}
\caption{The dispersion relation of a non-free operator, drawn over a Brillouin zone.}\label{F:ndNonfreeDisp}
\end{center}
\eef

The following remark is often very useful:
\bl\label{L:dense}
Let $S$ be a dense subset of the Brillouin zone $\B$. Then
\be\label{E:dense}
\sigma(H)=\overline{\bigcup\limits_{k\in S}\sigma(H(k))}.
\ee
\el
In other words, if for some reason one wants to avoid some exceptional values of $k$ (e.g., $k=0$ causes troubles for the Maxwell operator), one can avoid them by choosing an appropriate dense subset of quasi-momenta.

We can now define the spectral bands of the operator $H$:

\bd\label{D:ndspbands}
The segment
\be\label{E:ndSpBand}
I_j:=\mathop{range}(\lambda_j)
=\{\lambda\in\R\,|\,\exists k\in\R^n \mbox{ such that }\lambda=\lambda_j(k)\}
\ee
is the \textbf{$j$th band of the spectrum} $\sigma(H)$.
\ed

Due to $\G^*$-periodicity with respect to $k$, one can also write
\be\label{E:ndSpBand}
I_j
=\{\lambda\in\R\,|\,\exists k\in\B \mbox{ such that }\lambda=\lambda_j(k)\}.
\ee

\bc\label{C:sp=union}\indent
\begin{enumerate}
\item Each band $I_j$ is a finite closed interval, both of whose endpoints tend to infinity when $j\to\infty$.
\item The bands cover the whole spectrum:
\be\label{E:sp=union}
\sigma(H)=\bigcup\limits_{j\in\Z^+}I_j.
\ee
\item The \textbf{bands can}\footnote{And, as we will see, usually do.} \textbf{overlap}.
\end{enumerate}
\ec

As it was in 1D, it is easy to check that for the free operator $H=-\Delta$,  $B_H$ is the union of the paraboloid $\lambda=k^2$ and its $\G^*$-shifts.
One can deduce from this (which requires highly non-trivial work involving number theory (e.g., \cite{Skr_psim85,Parn_BS,ParnSob_Invent})) the following two theorems:
\bt\label{T:overlap}
\textbf{Consecutive spectral bands of the free operator overlap,} and thus the \textbf{spectrum of the free operator has no gaps}.
\et
This is in a stark contrast with 1D, where the bands do not overlap (touching in the worst case).

To quantify the overlap of the bands $I_j$, Skriganov (see e.g. \cite{Skr_psim85}) introduced the following continuous \textbf{overlap function} on $\R$:
\bd\label{D:overlap}
\be
\zeta(\lambda):=
\begin{cases}
\max\limits_j\max \{t\geq 0\,|\, [\lambda-t,\lambda+t]\subset I_j\}, \lambda\in\sigma(H)\\
0, \lambda\notin\sigma(H)
\end{cases}.
\ee
\ed
\bt\label{T:overlapquant}\indent
\begin{enumerate}
\item In dimension $n=2$, one has
\be
\lim\limits_{\lambda\to\infty}\zeta(\lambda)=\infty.
\ee
\item When $n=3$,
\be
\lim\sup_{\lambda\to\infty}\zeta(\lambda)\geq c>0.
\ee
\item In dimensions $n\geq 4$, the lower bounds for the overlap functions are given by negative powers of $\lambda$ (see for details and further references \cite{ParnSob_Invent}).
\end{enumerate}
\et

\subsection{Periodic operators as perturbations of the free ones.}\label{SS:perturb}

If one understands the free operator well, there is a temptation to consider the periodic potential as a perturbation term, at least at high energies $\lambda\to\infty$. This happens to be a highly valuable approach in many instances (e.g., in establishing the finiteness of the number of spectral gaps and various spectral asymptotics, topics addressed later in this text). However, it is also challenging and technically difficult. Even moderately comprehensive exposition alone might take a book (and in fact such books have been written by Karpeshina and Veliev \cite{Karp_LNM,Veliev_book}). I thus will just indicate some of the difficulties and send the reader to the original texts for the details of various techniques employed.

It has been noticed (e.g., \cite{FKTstable,FKTunstable,FriedlPerturb,FriedErratum}) that the Floquet eigenvalues $\lambda_j(k)$ can be split into the sets of ``nice'' \textbf{stable} and much more complicated \textbf{unstable} or \textbf{resonant} ones. The perturbation approach works easier in the stable region and requires a lot of ingenuity working with the unstable one (in particular, showing that the resonant values are sufficiently rare). The reader is directed to \cite{Karp_LNM,ParnSob_Invent} and references therein for technical details and further references.

Looking back at Theorem \ref{T:overlapquant}, one can expect the difficulties growing with the dimension, which indeed does happen. It is also much harder to handle the higher order perturbations, e.g. periodic magnetic potential terms of Schr\"odinger operator. Again, Parnovski and Sobolev's paper \cite{ParnSob_Invent} is a good reference to look into for contemporary results and techniques.

\subsection{Born-Karman approximation}
A very useful way of obtaining the dispersion relation (and thus the spectrum) of a periodic operator is due to Born and Von Karman (see, e.g., \cite{AshcroftMermin_solid}).

We will assume, as before, that $\G=\Z^n$ (although the approach works for any lattice).
\bt\label{T:BornKarm}
Consider for an integer $N$ the cube $Q_N:=[-N,N]^n\in\R^n$ and define the operator $H^{(N)}$ acting as the differential expression of $H$ on $Q_N$ with periodic boundary conditions (i.e., on the cube $Q_N$ folded onto a large torus). Then
\be\label{E:BornKarman}
\sigma(H)=\lim\limits_{N \to \infty}\sigma(H^{(N)})=\overline{\bigcup\limits_{N\in\Z^+}\sigma(H^{(N)})},
\ee
where the limit is taken with respect to the Hausdorff distance on any bounded part of the real axis.
\et
Indeed, let us denote by $\Q\B$ the subset of the Brillouin zone consisting of all quasi-momenta with components that are rational multiples of $\pi$. Then, it is a rather straightforward exercise to see that
$$
\bigcup\limits_{N\in\Z^+}\sigma(H^{(N)})=\bigcup\limits_{k\in\Q\B}\sigma(H(k)).
$$
Now, using Lemma \ref{L:dense}, one proves Theorem \ref{T:BornKarm}.

\br\label{R:spurious}
Another important observation is that this theorem is more than just the claim about Hausdorff convergence of spectra, or density of states convergence. Indeed, such claims hold also for other ``reasonable'' choices of boundary conditions (e.g., Dirichlet), or other choices of large domains instead of large cubes assembled from the fundamental unit cubes \cite{Shubin_ap}. In those cases, however, one can often see \textbf{spurious spectra} appearing outside $\sigma(H)$, but disappearing in the limit. In the Born-Karman case, though, this does not happen, since $\sigma(H^{(N)})\subset \sigma(H)$ for any $N$.
\er
\subsection{Analytic properties of the dispersion relation $B_H$}
We will explore now analytic properties of the dispersion relation of the operator $H$. Consider the complex Bloch variety
\be
B_{H,\C} \subset \C^{n+1}.
\ee

\bt\label{T:analBloch}(\cite[Lemma 8 of \$4]{Kuc_rms82},\cite[Theorem 4.4.2]{Kuc_floquet}, see also \cite{KnoTru_cmh90})
\indent
\begin{enumerate}
\item There exists an entire function $f(k,\lambda)$ on $\C^{n+1}$, such that\indent
    \begin{enumerate}
\item \be|f(k,\lambda)|\leq C_p e^{(|k|+|\lambda|)^p}\ee
for any $p>n$. (Similar statement holds for more general elliptic periodic operators, with the exponent $p$ depending on the dimension and order of the operator \cite{Kuc_floquet}.)
\item \be
B_{H,\C}=\{(k,\lambda)\in \C^{n+1}\,| \, f(k,\lambda)=0\}.
\ee
\end{enumerate}
\item If we only allow $p>n+1$ as the exponential order, one can make sure that $f$ is $\G^*$-periodic w.r.t. $k$.
\end{enumerate}
In particular, $B_{H,\C}$ is a \textbf{$\G^*$-periodic complex analytic sub-variety of $\C^{n+1}$ of co-dimension $1$}.

\et

An easier statement, without estimates on growth of the function $f$, can be obtained from the analytic Fredholm theory (e.g., \cite{ZaiKreKucPan_umn75,Kuc_floquet,Wil_jam78}).
The proof of theorem as stated uses the theory of \textbf{regularized determinants in Shatten-von Neumann classes}, presented, e.g., in \cite{GohbergKrein_nonselfadjoint,Simon_traceideals,Dunford_v2}.

\subsection{Floquet variety}

It is natural to reduce $B_H$ with respect to its $\G^*$-periodicity (i.e., using Floquet multipliers $z=e^{ik}$ instead of crystal momenta k). We thus introduce the following (not commonly adopted) notion:

\bd\label{D:FloVar}
The {\bf Floquet variety $\F_H$ of $H$} is defined as follows:
\be\label{E:FloVar}
\F_H:= \{(z,\lambda)\in (\C^*)^n\times\C\, |\, z=e^{ik}=(e^{ik_1}, ... ,e^{ik_n}), \mbox{ with } (k,\lambda)\in B_{H,\C}\}.
\ee
Here $\C^*$ is the \textbf{punctured complex plane} $\C^*:=\C \setminus \{0\}$.
\ed
In other words, $(z,\lambda)\in\F_H$, iff there exists a non-zero Floquet solution of $Hu=\lambda u$ with the Floquet multiplier $z$.

\subsection{Non-algebraicity}
As we have seen already, in the free case, $B_H$ is the union of $\G^*$-shifts of a single paraboloid. Thus, we get
\bp\label{P:AlgIrred}
In the free case,\indent
\begin{enumerate}
\item All irreducible components of the Bloch variety $B_H$ are algebraic (shifted paraboloids). In particular, the Bloch variety $B_H$ is irreducible modulo $\G^*$-shifts.
\item The Floquet variety $\F_H$ is irreducible.
\end{enumerate}
\ep
It is thus interesting to ask what to expect in terms of algebraicity and irreducibility in the non-free periodic case.

In general, the irreducible components of neither Bloch, nor Floquet variety are algebraic (Feldman, Kn\"orrer, Trubowitz \cite{FKT_NoLame}). There exist, however, examples of
  \textbf{non-selfadjoint periodic operators $H$ with algebraic components of $B_H$} (see \cite[Theorem 11]{kuch_floquet1982} and \cite[Theorem 4.1.2]{Kuc_floquet}).

\textbf{In discrete problems}, the \textbf{Floquet varieties are algebraic} \cite{Kuc_incol89,BerKuc_book}, since the corresponding operators $H(z)$ are finite Laurent series with respect to $z$: $$\sum\limits_{j\in \Z^n} H_j z^j,$$ where the sum is finite and $H_j$ are matrices of a finite size.

An impressive detailed study of the structure the Bloch variety for discrete periodic Schr\"odinger operators on $\Z^2$ was done by Gieseker, Kn\"orrer and Trubowitz \cite{GieKnoTru_cm91,Gieseker_fermi}.

Another important question is about irreducibility of the Bloch variety in higher dimensions, which is addressed in the next sub-section.

\subsection{Irreducibility}
It is conjectured that the analog of the statement (2) of Proposition \ref{P:AlgIrred} holds in higher dimensions:

\bcon\label{Con:irred}The Bloch variety of the self-adjoint periodic Schr\"odinger operator $H=-\Delta+V(x)$ is irreducible modulo $\G^*$.

A stronger version conjectures the same for more general periodic self-adjoint second order elliptic operators.
\econ

Proving this conjecture seems to be extremely hard. It has been proven in dimension $2$ by a tour de force in the paper \cite{KnoTru_cmh90} by Kn\"orrer and Trubowitz, as well as in the $2D$ discrete case by them and Gieseker \cite{GieKnoTru_cm91,Gieseker_fermi}.

A much weaker conjecture formulated below is already non-trivial:
\bcon\label{Con:noflat}The Bloch variety $B_H$ for a periodic elliptic \textbf{second order} operator $H$ with sufficiently ``decent'' coefficients does not have any flat components $\lambda$=\emph{const}.
\econ
We will see later on, how important is the resolution of this conjecture for the spectral theory. That is why it has attracted attention of many mathematicians and has been proven in many cases (albeit still not in its full generality), see Section \ref{SS:ac,pp,sing}.

We provide here an easy to complete sketch of the proof of the following well known result, due to Thomas \cite{Tho_cmp73}, although he did not formulate the statement in this form (see also \cite[Section VIII.16]{ReedSimon_v14}):
\bt\label{T:Noflat}\cite{Tho_cmp73}
Let $V\in\L_\infty(\R^n)$ be periodic. Then the Bloch variety of the Schr\"odinger operator $-\Delta+V(x)$ has no flat components $\lambda=\lambda_0$. (Notice that it is not assumed that the potential $V$ is real.)
\et

\begin{proof}
Suppose that this is incorrect, i.e. there exists $\lambda_0\in\C$ such that the equation $-\Delta(k)u+Vu=\lambda_0u$ has a non-trivial 1-periodic solution $u$ for any $k\in\C^n$. Absorbing the spectral parameter into the potential, we can assume that $\lambda_0=0$. Thus, we conclude that
$$
\Delta(k)u=Vu\mbox{ has a non-trivial solution $u$ for any }k\in\C^n.
$$
Let $M:=\|V\|_{L_\infty}$. Then $\|Vu\|_{L^2(\T)}\leq M \|u\|_{L^2(\T)}$ and thus
$$\|\Delta(k)u \|_{L^2(\T)}\leq M \|u\|_{L^2(\T)}$$ for any $k$ (with a non-trivial $u$ depending on $k$).

Let us now pick quasimomentum with a large imaginary part and judiciously chosen real part. Namely, assuming that $\G=\Z^n$ (analogous consideration works for any lattice), let
$$k=(\pi+i2a, 0, ... ,0), \, a\in\R.$$
Expansion of the periodic function $u$ into Fourier series easily leads to the estimate
$$\|\Delta(k)u \|_{L^2(\T)}\geq |a|\|u \|_{L^2(\T)}.$$
If $a>M$, we get a contradiction, unless $\|u\|=0$, which we assumed does not happen.
\end{proof}

The assumption of boundedness of the potential is way too strong (see, e.g. \cite[Section VIII.16]{ReedSimon_v14}).

In fact, a stronger statement holds:
\bt\label{T:FermiZero}
Under the same conditions as in Theorem \ref{T:Noflat}, for any $\lambda_0\in\R$, the level set of the dispersion relation (\textbf{isoenergy surface}, or \textbf{Fermi surface}, see Section \ref{SS:Fermi})
\be
F_{\lambda_0}=\{k\in\R^n\,|\,H(k)u=\lambda_0u\mbox{ has a non-zero solution }\}
\ee
has measure zero in $\R^n$.
\et

We will see later that this leads to the absolute continuity of the spectrum of periodic Schr\"odinger operators, which was the main goal in \cite{Tho_cmp73}. Due to this relation, the statements of Theorems \ref{T:Noflat} and \ref{T:FermiZero} have been extended in the last three decades to a large variety of scalar and matrix periodic operators of mathematical physics (e.g., Pauli, Dirac, magnetic Schr\"odinger, Maxwell, and elasticity operators, see Section \ref{SS:ac,pp,sing} and references therein), although the proof of the full Conjecture \ref{Con:noflat}
remains elusive.
\br\label{R:Dyakin}\indent
\begin{enumerate}
\item The consideration of the proof of Theorem \ref{T:Noflat} can be carried over to a variety of (probably not too interesting) periodic operators $L(D)+V(x)$, where $L(D)$ is a constant coefficient operator \cite[p. 28 and references therein]{kuch_floquet1982}.
\item An often overlooked interesting observation was made by Dyakin and Petruhnovskii \cite{DyakinPetrDiscr,DyakinPetrFermi} that the Fermi surface cannot contain not only flat parts, but also open pieces of some non-flat rational curves.
\end{enumerate}
\er
We need first the following auxiliary statement:
\bl\label{L:measurezero}
Let $a(k)$ be a real analytic function on (a connected open domain of) $\R^n$. If the measure of the zero set of $a$ is positive, then the function vanishes identically.
\el
\br This claim has several elementary proofs, but for some reason it is hard to find it even stated in most texts on real analytic geometry (\cite{KrantzRA} is an exception). Due to the text size limitations, let me just indicate three approaches, each of which can be easily completed by a graduate student: using Fubini's theorem (as suggested in \cite[Corollary 5.2]{ShubinBerezin_schrod}); by induction with respect to the dimension (suggested in \cite[begginning of Section 4.1]{KrantzRA}); by covering the set with a countable set of smooth submanifolds, obtained using implicit function theorem \cite{MitAnal}. It can also be derived from the Weierstrass Preparatory Theorem (e.g., \cite{GunnRossi_complex}). It is also a direct corollary of the Lojasiewicz's result on existence of Whitney stratification of real analytic sets \cite{LojasStrat,LojasBook}, as well as of the desingularization theorems by Bierstone and Milman \cite{BierstMilman,BierstMilman91}, which seems to be quite an overkill in both cases. Another non-trivial derivation can be found in Federer's book \cite[Section 3.1.24, p. 240]{Federer}.
\er

Now, Theorem  \ref{T:FermiZero} becomes a simple corollary of Theorem \ref{T:Noflat}. Indeed, the level set of the dispersion relation is the set of real zeros of an entire function $f(k,\lambda_0)$ (see Theorem \ref{T:analBloch} for its definition). Thus, if this set has a positive measure, then, according to Lemma \ref{L:measurezero}, the function vanishes identically, and thus $f(k,\lambda_0)=0$ for all $k\in\C^n$. This is impossible according to Theorem \ref{T:Noflat}.

One should notice that for periodic elliptic operators of orders higher than 2 neither of these conjectures holds true. E.g., there are examples of such operators of the 4th order whose Bloch variety is reducible, and moreover, has a flat irreducible component \cite[pp. 135--136]{Kuc_floquet}.

Let us address another issue related to irreducibility.
We have seen that in the free case any non-zero $\G^*$-shift of an irreducible component $\Sigma$ of $B_H$ produces another component, different from $\Sigma$. In other words, irreducible components are not invariant with respect to any non-zero $\G^*$-shifts. It seems that in presence of a generic periodic potential the situation is different.
\bcon\label{Con:CompShifts}
For a generic potential, unless an irreducible component $\Sigma\subset B_H$ is algebraic, there exists a non-zero $\G^*$-shift that leaves it invariant.
\econ
If this is correct,
then perturbation of the free operator by a periodic potential not only deforms the paraboloids, but also glues them together.

\subsection{Extrema}
Since, according to (\ref{E:ndspect}) -- (\ref{E:ndspect2}) the edges of the spectrum occur at extrema of the band functions, studying these extrema is an important task.

\subsubsection{Location}
In $1D$, the extrema of the dispersion relation $\lambda(k)$ occur only at $k=0$ and $k=\pi$ in
the Brillouin zone, due to the monotonicity (Lemma \ref{L:monot}) of band functions on the reduced Brillouin zone.

Let us try to look at the points $k=0$ and $k=\pi$ in a different light. The reciprocal lattice $\G^*$ is invariant with respect to the symmetry group generated by its shifts and ``time reversion'' $k\mapsto -k$. One discovers then that the points $k=0,\pi$ are exactly the fixed points of some of these symmetries inside of the reduced Brillouin zone.

In higher dimensions, the symmetry group can be larger \cite{Wigner}. If one looks at the free case, one discovers that critical points of these overlapping paraboloids do occur at some symmetry points as well. Most computations for periodic Schr\"odinger and Maxwell operators also showed this effect. This has led to the following

{\bf formerly popular belief:}   \emph{For $-\Delta+V(x)$, the extrema of dispersion relation
are attained at fixed points $k\in \B$ of some symmetries (probably at the highest symmetry points).
I.e., computing the dispersion only around the symmetry points gives the correct spectrum as a set.}

No monotonicity (or any other) reason for this claim exists.
Some numerical evidence against it has been around, but not
widely known and/or believed. \textbf{This claim was discussed and disproved} analytically in \cite{HarKucSob_jpa07,ExnKucWin_jpa10} (see also simpler waveguide examples in papers \cite{BorPankrGuideExtr,BorPankrGuideExtr2} by Borisov and Pankrashkin and a physics paper \cite{Dangers}). It is also discussed in \cite{HarKucSob_jpa07} why in practice the extrema are indeed often located at the symmetry points.

\subsubsection{Generic structure of spectral edges}
\indent

After discussing the possible locations of the extrema, we move to the probably more important question of their structure.
The following complications can (and do) occur at a spectral edge value $\lambda=c$:
\begin{description}
\item[A] The extremal value $c$ is attained by more than one band function.
\item[B] The extremum $c$ of a single band function $\lambda_j$ is non-isolated (i.e., a ``Mexican hat'' type picture occurs).
\item[C] The extremum is isolated, but degenerate (i.e., the Hessian of $\lambda_j(k)$ at the extremum point is degenerate).
\end{description}
Although all of the above can occur (e.g., \cite{ShterExample,FilKach}), the general belief is that \textbf{generically} they do not (see, e.g., Colin de Verdiere \cite{Col_msmf91}).

\begin{conjecture}\label{Con:generic} Generically (with respect to the potentials and other free parameters of the operator), the extrema of band functions
\begin{enumerate}
\item are attained by a single band;
\item are isolated;
\item are non-degenerate, i.e. have non-degenerate Hessians.
\end{enumerate}
\end{conjecture}
In other words, one conjectures that generically near a spectral edge the dispersion relation looks like a single parabolic shape, i.e. resembles the dispersion relation at the bottom of the spectrum of $-\Delta$.
As we will see later (Section \ref{S:Threshold}), then various analogs of the properties of the Laplacian would be applicable to a generic periodic elliptic operator.

Why would one conjecture this? Well, existence of a degenerate extremum of the dispersion relation is an analytic equality type restriction. Then it is natural to think that it should either hold almost never, or for (almost) all operators. It is hard to believe that such a restriction for all periodic potentials exists.

The common idea is that generically, the dispersion relation probably behaves like the spectrum of a ``generic'' family of self-adjoint matrices. This has been conjectured in various (explicit or implicit) forms by several authors (e.g., by Avron and Simon, Colin de Verdiere, and Novikov \cite{Col_msmf91,AvrSim_ap78,NovTypical,Novikov_viniti}).

Although the validity of the conditions (1)-(3) of the Conjecture \ref{Con:generic} is often assumed in mathematics and physics literature (it is involved with the definition of effective masses in solid state physics, homogenization, Green's function asymptotics, Liouville type theorems, impurity spectra in lacunas, Anderson localization, etc., see Section \ref{S:Threshold}), the conjecture remains largely unproven.

Let us mention what is known in this regard.

The following classical result by Kirsch and Simon \cite{KirSim_jfa87} establishes the conjecture (in a stronger form) at the bottom of the spectrum of a Schr\"odinger operator with periodic electric potential:
\begin{theorem}\cite{KirSim_jfa87}\label{T:K-S} Let
$$H=-\Delta +V(x)$$
be a periodic Schr\"odinger operator in $\R^n$. Then the bottom of the spectrum of $H$ is attained by the non-degenerate minimum at $k=0$ of the lowest eigenfunction $\lambda_1$ only.
\end{theorem}
An analogous result was obtained by Birman and Suslina for the periodic $2D$ Pauli operator \cite{BirSusPauli}. An improvement to Theorem \ref{T:K-S}, under additional symmetry assumption on the potential, was obtained by Helffer and T.~Hoffmann-Ostenhof \cite{HelfHofOst_reflsymmetries}. Regretfully, the arguments of \cite{KirSim_jfa87,BirSusPauli} are not applicable in the case of an internal spectral edge.

The conjecture was not only stated, but also proven for ``small'' potentials in $2D$ by Colin de Verdiere \cite{Col_msmf91}:

\bt \cite{Col_msmf91} \label{T:ColinGeneric}
For given lattice $\G\subset \R^2$ and natural number $N$, there exists a residual set $R$ in a neighborhood of zero in $C^\infty(\T,\R)$, such that for any potential $V\in R$ and all $j\leq N$, the Floquet eigenvalues $\lambda_j(k)$ are simple and $\lambda_j(k)$ is a Morse function on the torus $\T$.
\et

The following partial result was proven by Klopp and Ralston \cite{KloRal_maa00}:
\begin{theorem} \cite{KloRal_maa00} \label{T:KloppRals}
The claim (1) of Conjecture \ref{Con:generic} holds for a generic Schr\"odinger operator with periodic electric potential.
\end{theorem}

The second claim of the Conjecture was recently proven for Schr\"odinger operators in $2D$ by Filonov and Kachkovskii \cite{FilKach}.

It was shown by Shterenberg \cite{ShterLOwerEdgeMagn,ShterExample} that presence of a magnetic potential can create degeneracies even at the bottom of the spectrum.

The conjecture makes sense also \textbf{in the discrete case} and was \textbf{proven for periodic Schr\"odinger operators on $\Z^2$ by Gieseker, Kn\"orrer and Trubowitz} (see the overview \cite{GieKnoTru_cm91} for the results and exact references).

For more general $Z^2$-periodic graphs there is the following partial result:
\begin{proposition}\cite{DoKucSott}
Let $H$ be the periodic Laplace-Beltrami type operator\footnote{See \cite{BerKuc_book,Chung_spectralgraph} for the explicit form of such operators.} on a $\Z^2$-periodic graph $\G$, having just two vertices (atoms) per a fundamental domain, with all possible edges between adjacent copies of fundamental domain being allowed. Then the set of parameters (vertex and edge weights) for which the dispersion relation has a degenerate extremum is a (semi-)algebraic subset of co-dimension $1$ in the space of all parameters.
\end{proposition}
However, as an example by Filonov \cite{FilKach} of a stable nonisolated extremum shows, one has to be careful, allowing for a sufficiently large space of perturbation parameters (analogs of potentials).

In principle, if the conjecture were proven for the discrete case, one could attempt to bootstrap to carry the result over to the continuous case (such a procedure was used for a different purpose in \cite{HarKucSob_jpa07}). A direct approach to the continuous case, skipping the discrete one, would be desirable.

Allowing change of the lattice of periods to a sub-lattice, possibility of removing degenerate edges by perturbation of the potential was shown in $2D$ by Parnovski and Shterenberg \cite{FilParnShter}.

\subsection{Dirac cones}\label{SS:DCones}

When two branches of the dispersion relation meet, they often form a conical junction point, called a Dirac cone or sometimes a ``diabolic point'' \cite{BerryDiab}. The former name reflects the fact that the conical structure resembles the one for the dispersion of a 2D massless Dirac equation \cite{KatsGraph}.
\bef[ht!]
\begin{center}
\includegraphics[scale=0.5]{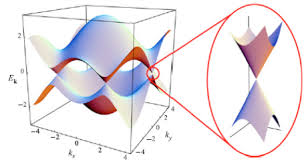}
\caption{Dirac cones for the tight-binding model of graphene.}
\end{center}\label{F:DCone}
\eef

Such conical structures are usually unstable under perturbation of parameters (e.g., the potential) of the operator. However, it has been noticed that if the structure has honeycomb lattice symmetry, preserved under perturbation, this protects the cone from splitting. Thus, in particular, such a cone mandatorily arises in the dispersion relation of  the famous \textbf{graphene}, which explains its amazing electric properties (see, e.g., \cite{KatsGraph}).

The observation of the mandatory appearance of Dirac cones in honeycomb-symmetric structures
\bef[ht!]
\begin{center}
\includegraphics[scale=0.3]{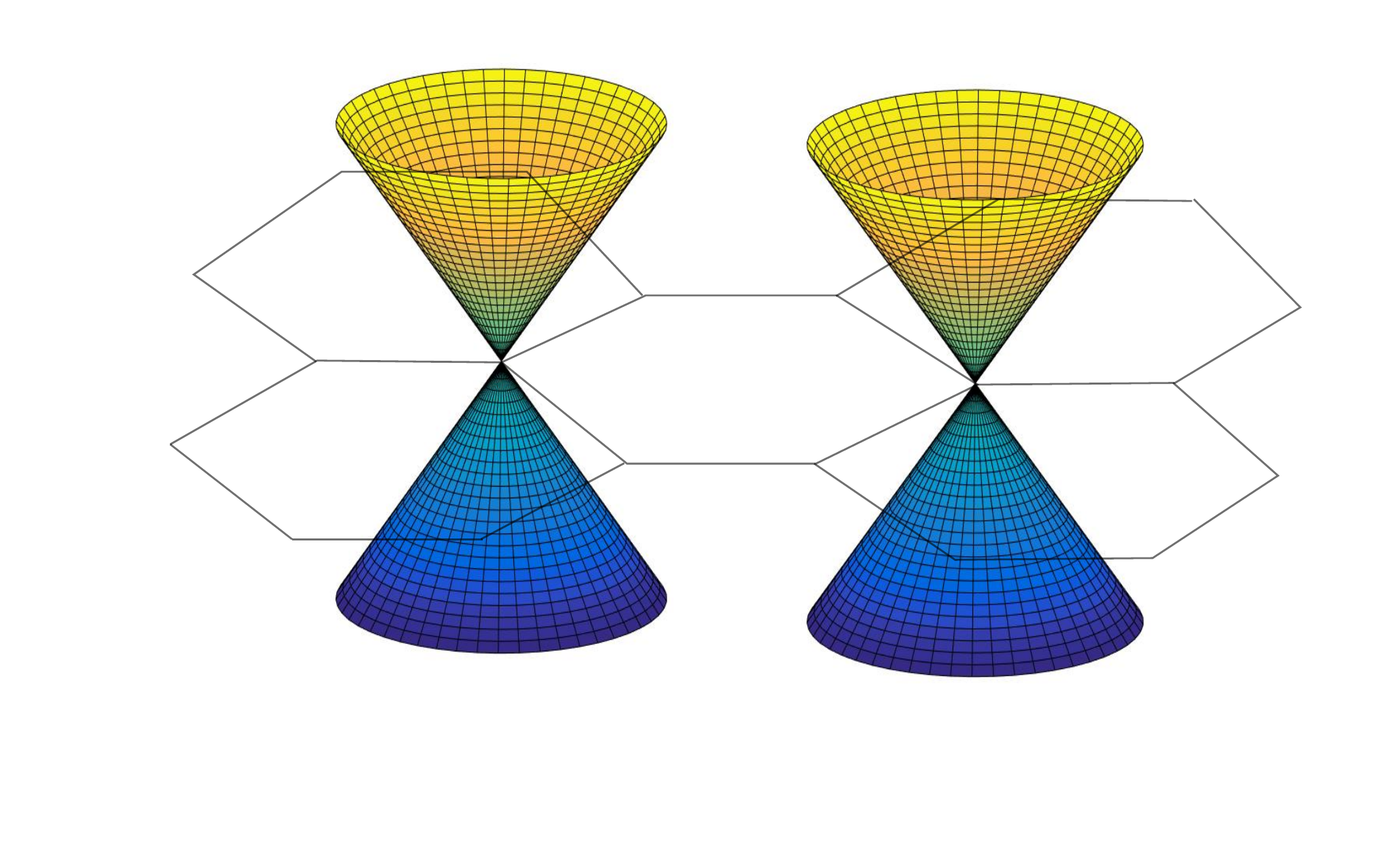}
\caption{Dirac cones for the honeycomb structure.}
\end{center}\label{F:graphenecones}
\eef
was first made (way before graphene came into play) for the tight binding model of the discrete honeycomb lattice by Wallace \cite{Wallace}, which provides an approximate picture of the graphene's first two dispersion bands. An infinite-band quantum graph model, which has an infinite-dimensional freedom of choosing honeycomb--symmetric potentials was considered in \cite{KucPos_cmp07}, where detailed structure of the spectrum and dispersion relation, including in particular presence of Dirac cones, was described\footnote{The article  \cite{KucPos_cmp07} also contained the detailed spectral analysis of quantum graph models of all possible types of carbon nanotubes.}. For 2D Schr\"odinger operators with honeycomb-symmetric potentials, existence and stability of Dirac cones was established, explained, and studied by Grushin \cite{Grushin}, Fefferman and Weinstein \cite{FefWei_pre12,FefWein1Dtopol,FefWeinEdgeHoney,FefWeinBifEdge,FefWeiPackets}, and Berkolaiko and Comech \cite{BerkComech}. The Schr\"odinger operator with honeycomb lattice of point scatterers was considered in \cite{PointscattMLee}. Carbon nanotubes as folded sheets of graphene have been studied in various papers, see e.g. \cite{KorLob_ahp07,Do,KucPos_cmp07} and references therein.

One can say, however, that the story does not end here, since presence of stable Dirac cones (including very interesting non-isotropic ones) has been observed for more general 2D (\textbf{graphyne}) structures, which lack honeycomb symmetry. See, e.g. \cite{graphyneEI} for physics discussion and \cite{Do,DoKuc} for quantum graph studies of a simplest graphyne. So far, there is no complete understanding of this phenomenon (analogous to the one provided in \cite{Grushin,FefWei_pre12,BerkComech} for the honeycomb case).

\subsection{Fermi surfaces}\label{SS:Fermi}
The notion of the Fermi surface is among the central ones in the solid state theory, as one can see from any solid state textbook, e.g. \cite{AshcroftMermin_solid,Ziman_solids,Callaway_band,Kittel_solidstate}. There are plenty of books and databases dedicated to Fermi surfaces of various crystals, e.g.
\cite{Cracknell_fermi,CracknellWong_fermi,Fermi_database,Ziman_Fermi,Ziman_solids,TheFermiSurface,LocalGeomFermi}. The reason is that many physical properties of, e.g., metals depend upon the geometry of its Fermi surface. We will try to show that questions about the analytic geometry of Fermi surfaces are intimately related to some basic spectral and other properties of the relevant operator. Answering these questions is usually extremely hard (see, e.g., the Gieseker, Kn\"orrer, and Trubowitz' book \cite{Gieseker_fermi}, and a shorter survey \cite{GieKnoTru_cm91}, devoted solely to the case of a discrete periodic Schr\"odinger operator on $\Z^2$).

To define a Fermi surface, let us consider the (real or complex) dispersion relation $B_H$ as the graph of the multiple valued function $$k\in \C^n \mapsto \sigma(H(k)).$$

\begin{definition}\label{D:Fermi} The \textbf{Fermi surface}\footnote{Also called \textbf{isoenergy surface}.} $F_{\lambda,H}$ of $H$ at a scalar value $\lambda$ is the $\lambda$-level set
of the dispersion relation. I.e.,
\begin{description}
\item[Complex Fermi surface is]
$$
\begin{array}{c}
F_{\lambda,H,\C}:=\{k\in\C^n \,|\, (k,\lambda)\in B_{H,\C} \}\\
=\{k\in\C^n \,|\, H(k)u=\lambda u \mbox{ has a non-zero solution}\}.
\end{array}
$$
\item[Real Fermi surface for a real $\lambda$ is]
$$
\begin{array}{c}
F_{\lambda,H}:=\{k\in\R^n \,|\, (k,\lambda)\in B_H \}\\
=\{k\in\R^n \,|\, H(k)u=\lambda u \mbox{ has a non-zero solution}\}.
\end{array}
$$
\end{description}
\end{definition}
An obvious statement is:
\begin{lemma}
The Fermi surface is $\G^*$-periodic.
\end{lemma}
The following result follows from its analog for the Bloch variety $B_H$:
\begin{theorem} $F_{\lambda,H}$ is the zero set of an entire function $f(k)$ of the same exponential order as for $B_H$ (see Theorem \ref{T:analBloch}).
\end{theorem}
The following result, which holds in a very general setting of elliptic periodic operators (compare with Theorem \ref{T:FermiZero}) is useful:
\bt\label{T:FermiZero2}
At any energy level $\lambda\in\R$, the real Fermi surface either has measure zero in $\R^n$, or it coincides with the whole $\R^n$ (and thus the complex Fermi surface at this level is $\C^n$).
\et

\begin{remark}\indent
\begin{enumerate}
\item In physics, the name Fermi surface is used only for a specific real level of the energy $\lambda$, the so called \textbf{Fermi level} $\lambda_F$ \cite{AshcroftMermin_solid,Ziman_solids,Callaway_band,Kittel_solidstate}. It is convenient for us, though, to consider Fermi surfaces at arbitrary levels.
\item The definition of the Fermi surface seems to be rather innocuous. Just to impress on you its possible complexity, Fig. \ref{F:Fermi_niobium} presents the Fermi surface of Niobium (Nb).
\bef[ht!]\begin{center}
 \includegraphics[scale=1.1]{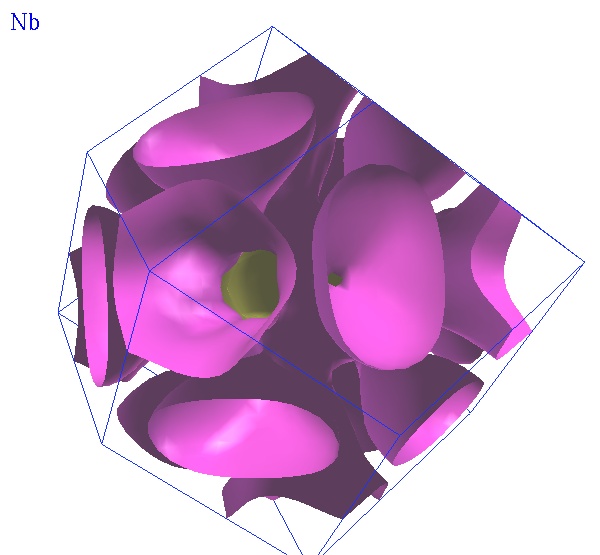}
 \caption{The Fermi surface of Niobium (Nb). The picture is borrowed from the URL \cite{Fermi_database}, where one can find many more (often more complex) pictures of Fermi surfaces.}\label{F:Fermi_niobium}
\end{center}
\eef
\end{enumerate}
\end{remark}
\begin{conjecture}\label{Co:FermiIrred}
The Fermi surface for $H=-\Delta+V(x)$ is irreducible (modulo $\G^*$) for $\lambda \in \R$, possibly except for a discrete set of values of $\lambda$.
\end{conjecture}
This conjecture has been proven by Gieseker, Kn\"orrer, and Trubowitz for discrete Schr\"odinger operator on $\Z^2$ \cite{Gieseker_fermi,GieKnoTru_cm91} and in the case of continuous Schr\"odinger operator with periodic electric potential in dimensions $n=2,3$ and separable potential $V(x)=\sum V_j(x_j)$, as well as in $3D$ for potential
$U(x_1,x_2)+V(x_3)$, where the axis $x_j$ are oriented along the basis vectors of $\G$ \cite{KucVai_cpde00}.

As we will see in Section \ref{SS:Imp}, the ``esoteric'' question of irreducibility of the Fermi surface is closely related to the question of absence of impurity eigenvalues embedded into the continuous spectrum.

\subsection{Bloch bundles}\label{SS:BlochBundle}
Let $S\subset \sigma(H)$ be a subset consisting of $m$ spectral bands and surrounded by spectral gaps. We will call such a subset a \textbf{composite band} (see Fig. \ref{F:composite}).
\bef[ht!] \begin{center}
\includegraphics[scale=0.6]{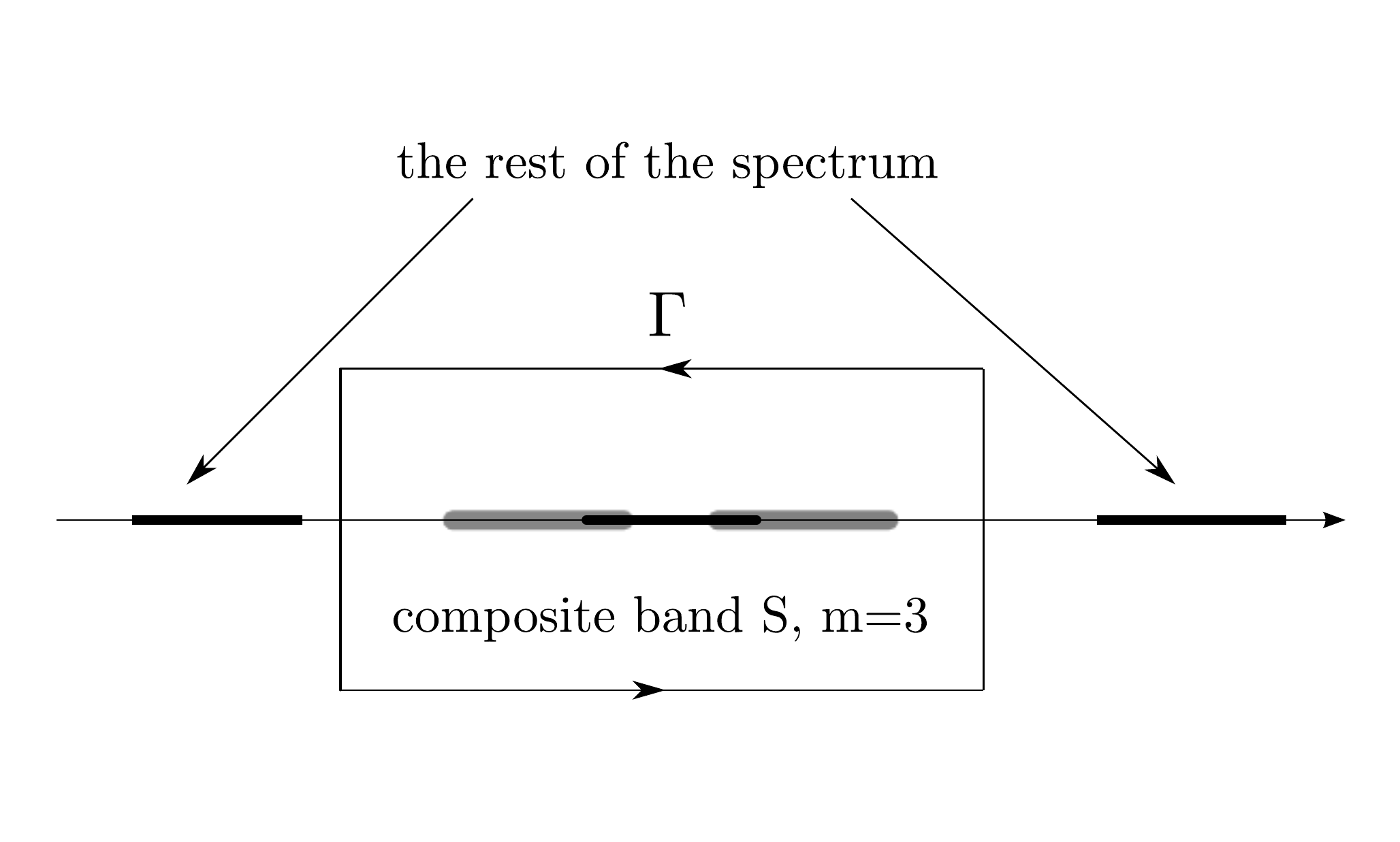}
\caption{A composite band composed of three spectral bands.}\label{F:composite}
\end{center}
\eef
Let us surround $S$ with a contour $\G$ (Fig. \ref{F:composite})  and introduce the corresponding $m$-dimensional spectral projector for $H(k)$:
$$
P(k):=\frac{1}{2\pi i}\oint_\G (\zeta - H(k))^{-1} d\zeta.
$$
Projector $P(k)$ depends analytically on $k$ in a complex neighborhood of $\T^*$ in $\C^n$.
Its range forms the so called \textbf{Bloch bundle} over this neighborhood,  which corresponds to the composite band $S$.
Sections of this bundle form the invariant subspace for $H$ that corresponds to $S\subset\sigma(H)$.

This bundle over the torus $\T^*$ can be topologically non-trivial (e.g., in presence of magnetic potential (Thouless \cite{Thouless}) or in the case of topological insulators \cite{BernTopIns}).

As we have seen, the bundle is defined and analytic in a complex neighborhood of the torus. We are sometimes interested in its analytic triviality. The question arises whether there are additional analytic obstructions to this. The following result is a simple incarnation of the so called Oka's principle \cite{GunnRossi_complex,GrauRemm_stein}:
\bt (e.g., \cite{KuchWannier}) Topological triviality of the Bloch bundle over $\T^*$ is equivalent to its analytic triviality in a complex neighborhood.
\et
Indeed, there is a neighborhood of the torus that is the product of 1D complex domains. Then it is a Stein manifold \cite{GrauRemm_stein}, and the above result is a part of the famous Grauert's theorem \cite{Grauert} for such manifolds.

A variety of results on sufficient conditions of triviality of Bloch bundles is known, see e.g. \cite{PanatiBlochBundle,KuchWannier} and references therein (see also \cite{ZaiKreKucPan_umn75} for the survey of techniques of Banach bundles).

\subsection{Analyticity in the space of parameters}\label{SS:anal_param}
An often very useful thing to do is to add to the spectral parameter $\lambda$ and quasimomenta $k$ an infinite-dimensional space $P$ of parameters of the periodic operator (i.e., an appropriate Banach space of periodic electric and/or magnetic potentials). Then the Fermi and Bloch varieties can be defined in this extended space. When the space of parameters is defined properly, it is usually straightforward to prove the following:

\begin{meta}\label{MT:anal_param} Bloch and Fermi surfaces are analytic subsets in the enlarged space $\C^{n+1}\times P$ (respectively $\C^{n}\times P$).
\end{meta}
\subsection{Inverse problems}\label{SS:inverse}

Inverse spectral problems are known to arise frequently and have been studied a lot for periodic ODEs (e.g., \cite{Nov_sol84,AblSolitons,DubKrich_book,McKean_Trub,LaxKdV,GarnTrub,GarnTrub2} and references therein).

The periodic PDEs in this regard are much trickier and there are not that many results available. We look here at a few examples.

\subsubsection{Borg's theorem}
The famous \textbf{Borg's uniqueness theorem} for the Hill operator $-\dfrac{d^2}{dx^2}+q(x)$ with periodic potential $q$ says:

\bt\cite{Borg}\label{T:borg1D}
The following claims are equivalent:
\begin{enumerate}
\item The potential is constant.
\item There are no spectral gaps.
\item \cite{AvrSim_ap78} There exists an entire function, whose graph $\lambda=f(k)$ belongs to the dispersion relation.
\end{enumerate}
\et
The equivalence $(1) \, \Leftrightarrow \, (2)$ fails miserably in dimensions $n>1$! In fact, as we will see later, for any sufficiently small bounded periodic potential $V(x)$ in $\R^2$ and $\R^3$, the Schr\"odinger operator $-\Delta+V(x)$ has no gaps in the spectrum. Moreover, when the gaps do appear, there are only finitely many of them. Thus, they clearly carry insufficient information for recovery of the potential. Spectral gaps in higher dimensions do not provide nearly as much information as in the ODE case. However, the equivalence $(1) \, \Leftrightarrow \, (3)$  probably still holds:
\begin{conjecture}\label{Con:Borg} The following claims are equivalent:
\begin{enumerate}
\item The potential is constant.
\item There
exists an entire function, whose graph $\lambda=f(k)$ belongs to the dispersion relation.
\end{enumerate}
\end{conjecture}

In $2D$, the conjecture was proven by Kn\"orrer and Trubowitz \cite{KnoTru_cmh90}.

\subsubsection{Floquet rigidity}
Let us consider the Schr\"odinger operator $H=-\Delta +V(x)$ with the potential periodic with respect to a lattice $\G$ and the corresponding Bloch operators $H(k)$, which have, as we know, discrete spectra.

\bd\label{D:PDEspectra}\indent
\begin{itemize}
\item The \textbf{periodic spectrum} is defined as follows:
\be
\sigma_0(H):=\sigma(H(0))=\mbox{ spectrum of $H$ on  $\G$-periodic functions.}
\ee
\item The \textbf{$k$-spectrum} is defined as follows:
\be
\begin{array}{c}
\sigma_k(H):=\sigma(H(k)) \\
=\mbox{ spectrum on Bloch functions with quasimomentum $k$.}
\end{array}
\ee
\end{itemize}
Clearly, the union over all $k$ of the graphs of these spectra forms the Bloch variety for $H$.
\ed

\bd\label{D:isospectral}\indent
\begin{itemize}
\item The $\G$-periodic potentials $V$ and $W$ are \textbf{isospectral}, if
\be
\sigma_0(-\Delta+V)=\sigma_0(-\Delta+W).
\ee
\item The $\G$-periodic potentials $V$ and $W$ are \textbf{Floquet isospectral}, if
\be
\sigma_k(-\Delta+V)=\sigma_k(-\Delta+W), \forall k\in\R^n.
\ee
\end{itemize}
\ed

One is interested in understanding when two potentials are isospectral, or even Floquet isospectral. These questions have been understood well in $1D$ (see, e.g., \cite{McKean_Trub}), while for $n\geq 2$ problems are far from being resolved.

There are simple examples of isospectral potentials. For instance, one observes
\bl\label{L:pm insosp}\indent
\begin{enumerate}
\item Potentials $V(x)$ and $W(x)=V(\pm x+a)$, for any $a\in\R^n$, are Floquet isospectral.
\item If the lattice $\G$ is invariant with respect to an orthogonal transformation $O$ of $\R^n$, then the potentials $V(x)$ and $V(Ox)$ are isospectral.
\end{enumerate}
\el
Imposing the following condition on the lattice $\G$, one eliminates the second option in the lemma above:
\bd\label{D:latt_condit}\indent
A lattice $\G$ has a \textbf{simple length spectrum}, if
\be\label{E:latt_condit}
\mbox{If }\g_1,\g_2\in\G\mbox{ and }|\g_1|=|\g_2| \mbox{, then }\g_2=\pm\g_1.
\ee
\ed

One other option is to use the well known results on $1D$ isospectrality. Indeed, let $v(s)$ and $w(s)$ be smooth $1$-periodic functions of one variable and $\delta$ be a vector such that
\be\label{E:delta_cond}
\{\delta\cdot\gamma\,|\, \g\in\G\}=\Z.
\ee
It is easy to establish \cite{EskRalTru_cpam84,EskRalTru_cpam84a} that \textbf{if $1D$ potentials $v$ and $w$ are isospectral, then the ``ridge'' $n$-dimensional potentials
$v(\delta\cdot x)$ and $w(\delta\cdot x)$ are isospectral.}
One wonders whether there are any other options for isospectrality (Floquet isospectrality)\footnote{One can easily achieve Floquet isospectrality by allowing complex potentials \cite[Theorem 4.1.2]{Kuc_floquet}.}.

Many significant results on multi-dimensional isospectrality were obtained by Eskin-Ralston-Trubowitz in \cite{EskRalTru_cpam84,EskRalTru_cpam84a} (see also an overview in \cite{ERT_contemp}), mostly under conditions of analyticity of the potentials and the simplicity of the length spectrum (\ref{E:latt_condit}). See also preceding and following works of various authors with additional results and related discussions in \cite{Bat_isosp,Waters,GuillTori,GordGuerKapp,GordKap,GordKap2,Kapp2,Kapp3,Kapp_isosp1,GuillS2,Moerb,Esk_edp89,Esk_cmp89,EskRalTru_cpam84}.

We present below a few results of \cite{EskRalTru_cpam84,EskRalTru_cpam84a}\footnote{The sophisticated techniques used in \cite{EskRalTru_cpam84,EskRalTru_cpam84a} to prove the results involved, in particular, the spectral invariants extracted from the heat and wave kernel asymptotics.}.

\bt\label{T:iso_floq}
Suppose that the length spectrum of $\G$ is simple and $\G$-periodic potentials $V$ and $W$ are analytic, then
 \begin{itemize}
 \item If $V$ and $W$ are isospectral, then they are Floquet isospectral.
 \item Let $k_0\in\R^n$ be such that $\cos (2\pi k_0\cdot\g)\neq 0$ for all $\g\in\G$. If $\sigma_{k_0}(-\Delta +V)=\sigma_{k_0}(-\Delta +W)$, then $V$ and $W$ are Floquet isospectral.
 \end{itemize}
\et
This result shows that (under appropriate conditions) the whole Floquet spectrum (i.e., the whole Bloch variety) has an overdetermined information, and the fiber of $B_H$ over an appropriate quasimomentum $k_0$ carries as much information as the whole $B_H$.

It is natural to look in the ``horizontal,'' i.e. along quasimomenta, direction. In particular,
\begin{center}
\textbf{does the knowledge of a single band function $\lambda_j(k)$ determine the whole dispersion relation (Bloch variety)?}
\end{center}
In fact, the Conjecture \ref{Con:irred} (proven in $1D$ and $2D$) implies, if correct, that \textbf{\emph{even any open part of a single branch  $\lambda_j(k)$ uniquely determines the dispersion relation.}}

Let now $V$ be a $C^\infty$ $\G$-periodic potential with the Fourier series expansion
\be\label{E:PotFourExp}
V(x)=\sum\limits_{\delta\in\G^*}a_\delta e^{i\delta\cdot x}.
\ee
\bd\label{D:reduc}
Let $\g\in\G$. The \textbf{reduced potential} $V_\g(x)$ is defined as follows:
\be\label{E:reduc}
V_\g(x)=\int\limits_0^1 V(x+s\g)ds = \sum\limits_{\delta\cdot \g=0}a_\delta e^{i\delta\cdot x}.
\ee
\ed
Another important observation in \cite{EskRalTru_cpam84,EskRalTru_cpam84a,ERT_contemp} concerns the relations between the Floquet spectra of the original and reduced potentials:

\bt\label{T:reduced}
For any $\g\in\G$, the Floquet spectrum $\sigma_k(-\Delta+V)$ determines uniquely $\sigma_k(-\Delta+V_\g)$ .
\et

Let now $\delta\in\G^*$ be such that there exists $\g_0\in\G$ for which $\delta\cdot\g_0=2\pi$. Then one can introduce the $1D$ \textbf{directional potential}
\be\label{E:V_d}
V_\delta(s):=\sum\limits_{n=-\infty}^\infty a_{n\delta}e^{ins}
\ee
\bt\label{T:direct}
The Floquet spectrum $\sigma_k(-\Delta+V)$ determines uniquely the Floquet spectra of the $1D$ spectral problems
\be\label{E:1Ddirect}
\left(-|\delta|^2\frac{d^2}{ds^2}+V_\delta(s)\right)u(s)=\lambda u(s)
\ee
for any directional potential $V_\delta$.
\et

The isospectrality of potentials $V(\pm x+a)$
suggests a question whether under some conditions one can prove that, modulo shifts, the number of isospectrality classes is finite. A variety of important results of this type can be found in the paper \cite{EskRalTru_cpam84a} by Eskin, Ralston, and Trubowitz. See also Gieseker, Kn\"orrer, and Trubowitz \cite{GieKnoTru_cm91,Gieseker_fermi} for inverse spectral results for discrete periodic operators on $\Z^2$.

An alternative approach to the inverse spectral problem for periodic potentials was provided by Veliev \cite[Ch. 4]{Veliev_book}.

\section{Spectral structure of periodic elliptic operators}\label{S:spectra}
As we have seen, both in $1D$ and in higher dimensions, periodic elliptic operators have band-gap structure of their spectra. There are, however, significant differences between ODE and PDE cases. We make the comparison in the table below:
\begin{center}
\begin{tabular}{|c|c|c|}
  \hline
  & $n=1$ & $n>1$ \\
  \hline
  band overlaps & none   & frequent \\ \hline
  gaps  & generically all open   & many gapless potentials \\ \hline
  no gaps & constant potential  & all small potentials \\ \hline
  free operator bands&touch to cover $\R^+$   & overlap to cover $\R^+$ \\
  \hline
\end{tabular}
\end{center}

The rest of this section is devoted to discussing various spectral properties of periodic operators, such as the band-gap structure of the spectrum, the qualitative structure of the spectrum, eigenfunction expansions, etc.

\subsection{Spectral gaps}\label{SS:gaps}

We start with considerations of spectral gaps: their existence, number, and ways of creating them.

\subsubsection{Existence of gaps}\label{SS:gaps_eist}

As we have discussed already, according to the Borg's theorem, in $1D$ every non-constant periodic potential creates spectral gaps.
This is manifestly incorrect in higher dimensions, e.g.:
\bt\label{T:nogaps}
If $n=2,3$ and the $L_\infty(\R^n)$-norm of a periodic potential $V(x)$ is sufficiently small, the Schr\"odinger operator $H=-\Delta+V(x)$ in $L^2(\R^n)$ has no spectral gaps.
\et
This is an immediate consequence of the first two claims of Theorem \ref{T:overlapquant}, if one considers $H$ as a perturbation of the free operator $-\Delta$.

\subsubsection{Maximal abelian coverings and Sunada's no gap conjecture}

Consider the standard covering $\R^n\mapsto \T \,  (=\R^n /\Z^n)$. Its deck group is $\Z^n$, which coincides
with $H_1(\T,\Z)$.
It can also be understood as the quotient of the universal cover of the torus by the commutator subgroup of the
 fundamental group $\pi_1(\T)$.
\bd\label{D:maxabel} The \textbf{maximal abelian covering} of a compact manifold $X$ is the covering $Y\mapsto X$
with the deck group $H_1(X,\Z)$, obtained as the quotient of the universal cover of $X$ by the commutator subgroup
of the fundamental group $\pi_1(X)$.
\ed
The following monotonicity result holds \cite{Sun_am85}:
\bt\label{T:sp_monot} Let $Y\mapsto X$ be a Riemannian covering with an amenable deck group. Then
$$\sigma(-\Delta_X)\subset \sigma(-\Delta_Y),$$
where $\Delta_M$ denotes the Laplace-Beltrami operator on a Riemannian manifold $M$.
\et
Thus, the spectrum of the Laplacian on the maximal abelian covering is the largest (and thus has the fewest gaps) among all abelian coverings of the same base. The following, still not proven, \textbf{no gap conjecture} has been formulated by Sunada \cite{Sun_jfsut90,Sun_incol08}:
\bcon\label{Con:nogap} Let $Y\mapsto X$ be the maximal abelian covering of a compact Riemannian manifold $X$ of a constant negative curvature. Then the spectrum of the Laplace-Beltrami operator $-\Delta_Y$ on $Y$ has no gaps.
\econ
Analogous conjecture has been formulated in the graph case, where the base graph $X$ is assumed to be regular (i.e., degrees of all its vertices are the same). The graph version of the conjecture was proven for all regular graphs of even degree \cite{HigShi_dga04} and for various examples of regular graphs of odd degree \cite{HigShi_dga04,kuch_K4}, e.g., the so called $K4$ graph $X$.

\subsubsection{Number of gaps. Bethe-Sommerfeld conjecture}\label{SS:BSC}
What about the possible number of spectral gaps, if they exist at all? In $1D$, the statement (3) of Theorem \ref{T:1Dspect} shows that generically the number of gaps is infinite \cite{KirSim_jfa87}.

The situation is quite different in dimensions $n>1$, as the following old {\bf Bethe-Sommerfeld Conjecture (BSC)} \cite{Bethe_elektronen} shows:
\bcon\label{con:BSC}
When $n>1$, there can only be a finite number of gaps in the
spectrum of any periodic Schr\"odinger operator
$$-\Delta+V(x)$$
in $\R^n$.
\econ
By now, after several decades of efforts, the conjecture has been proven in its full generality. The story started with the works by Popov and Skriganov\cite{PopSkrig} and Dahlberg and Trubowitz \cite{Dahlberg} for $n=2$ and by Skriganov \cite{Skr_im85} for $n=3$. Theorem \ref{T:overlapquant} shows that the issue should become much harder after that, and it does. For $n\geq 4$, the conjecture was established for the case of rational lattices \cite{Skr_psim85,SkrSob}. For $n=4$ the conjecture was proven for arbitrary lattices by Helffer and Mohamed \cite{HelfMochAsympt}. Finally, in Parnovski's paper \cite{Parn_BS}, the proof is provided in any dimension $n\geq 2$ and arbitrary lattice, under smoothness conditions on the potential (an alternative approach was offered by Veliev \cite{Veliev_book,BS_Vel07,BS_Vel88}). Predictably, the case of presence of magnetic potential turned out to be even harder. The result was established in the two-dimensional case by Mohamed \cite{MohElectromagn} and Karpeshina \cite{Karp2D}. A major advance both in techniques and results (allowing pseudo-differential lower order terms) was made by Parnovski and Sobolev in \cite{ParnSob_Invent} (this paper also contains a nice review and bibliography of the topic). Other related results on BSC (including, e.g., Maxwell operator) can be found in \cite{Shen_BS,BS_BarbParn,BS_elton,BS_Lapin,BS_sob07,ParnSvirid,ParnSobBS2000,ParnShter09,BS_Vorobets,MorozParn_density} and in the historical survey in \cite{ParnSob_Invent}.

\br\label{R:bsc}\indent
\begin{enumerate}
\item {\bf A stronger version of BSC}- for any second order periodic elliptic operator, is still not proven, except for
$H=(-\Delta)^m + \mbox{ lower order periodic terms}$
(see \cite[and references therein]{ParnSob_Invent}).

\item The Maxwell operator case is mostly unresolved, except in a very special case considered in the paper \cite{BS_Vorobets} by M.~Vorobets.

\item The validity of BSC for periodic waveguide systems has not been established and is sometimes questioned.

\item For periodic graphs/quantum graphs the statement of BSC does not hold \cite{BerKuc_book}.
\end{enumerate}
\er
\subsubsection{Gap creation}

Presence of gaps is necessary in many instances. E.g., they are responsible for properties of semi-conductors. Fortunately, such materials exist already in nature. The situation is different, for instance, when one tries to create the so called \textbf{photonic crystals}, where the operator of interest is Maxwell in a periodic medium and one needs to create such a medium with spectral gaps \cite{Joannopoulos_photocrystals,Kuch_pbg}. We know by now that periodicity, although leading to the band-gap structure of the spectrum, does not guarantee existence of gaps.
We thus address here briefly how spectral gaps can be ``engineered''.

Let us start with the Schr\"odinger operator $-\Delta+V(x)$. As we have just said, engineering gaps is not such a hot issue in this case. It also happens to be rather easy to create spectral gaps. Indeed, the following procedure can be easily made precise: creating a local potential well, one can create a bound state (eigenvalue) below the spectrum of the Laplacian. Now, repeating periodically copies of the well at large distances from each other will lead to spreading of this eigenvalue into a thin spectral band at a distance from the spectrum of the free operator, and thus creating a periodic medium with a spectral gap.

This potential well technique fails for Maxwell and some other operators, when local perturbation cannot lower the bottom of the spectrum. Another deficiency here is that the resulting gaps are rather hard to manipulate.

Another approach is creating very high contrast media, which succeeds in creating spectral gaps for periodic Maxwell operators \cite{FigKuc_siamjam96,FigKuc_siamjam96a,FigKuc_siamjam98,Fil_cmp03} as well as for Laplace-Beltrami operators on abelian coverings of compact Riemannian manifolds \cite[and references therein]{HempPost_gaps,Pos_jde03}. A two-scale homogenization approach has been developed by Zhikov  \cite{Zhikov_gaps}. An interesting series of papers on pre-designed gap opening was written by Khrabustovskyi and Khruslov \cite{Khrab_contgrol,KhrabAsymptPreassign,KhraKhru}.

The high contrast approach also has its deficiencies. First of all, it is still hard to manipulate the locations and sizes of spectral gaps. Besides, the high contrast frequently requires non-physical values of material parameters.

The moral is that in higher dimensions it often is hard to create and manipulate the spectral gaps arising due to Bragg scattering (i.e., due to the periodicity of the medium).

A different promising mechanism of opening \textbf{resonant gaps} exists. The author has learned about it first from Pavlov's work \cite{Pavlov}. The idea is that spreading infinitely many small identical resonators throughout the medium tends to create spectral gaps around the eigenvalues of the resonator. Regretfully, in the continuous case, this idea apparently has never been made precise. It has been implemented (as the so called \textbf{decoration procedure}), in its simplest incarnations in the discrete situation by Aizenman and Schenker \cite{SchAiz_lmp00}, as well as in the quantum graph case \cite{Kuc_jpa05,BerKuc_book}.
\bef
\begin{center}
\includegraphics[scale=0.4]{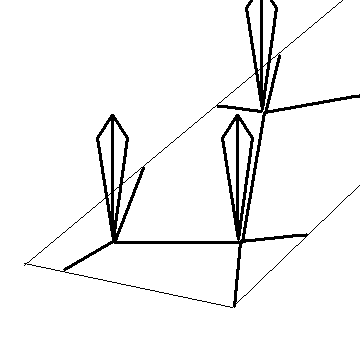}   \includegraphics[scale=0.4]{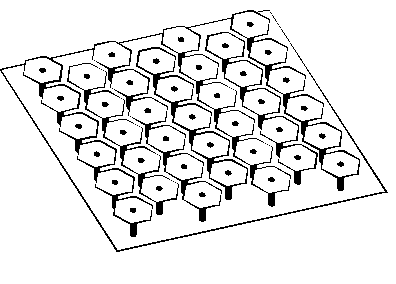}
\caption{A decorated graph (left). An antenna implementing the concept(right).}
\end{center}
\eef
In practice, a more involved ``spider decoration'' procedures are desirable (see discussion in \cite{BerKuc_book,Kuc_jpa05}), which are at an initial stage of development (see, e.g., \cite{Ong_diss,DoKucOng}).   One can compare this with the more elaborate \textbf{zig-zag} decoration procedure, used to create \textbf{expander graphs} (which boils down to controlling the size of the principal spectral gap) \cite{WidgZig_anm02}.

Resonant gaps can also be used to create ``slowing down light'' optical media (e.g., \cite{KucKun_em99,Yariv_crows99,Boyd_02,SmyshKuch}).

\subsection{Eigenfunction expansion}

Let us recall that
\be
\lambda_1(k)\leq \lambda_2(k)\leq \lambda_3(k) \leq \dots
 \ee
 is the sequence of eigenvalues of the operator $H(k)$, listed for $k\in\R^n$ (with their multiplicity) in a non-decreasing order.

The standard perturbation theory (e.g., \cite{Kato_perturbation}) implies that
\begin{center}
\textbf{each $\lambda_j(k)$ is a continuous, piece-wise analytic function of $k$.}
\end{center}
By definition, for each $\lambda_j(k)$ the operator $H$ has a generalized (since it is not square-summable) Bloch eigenfunction $\psi(x)$ with the quasi-momentum $k$ (i.e., $\psi(x)=e^{ik\cdot x} p(x)$ with a $\G$-periodic function $p(x)$):
\be\label{E:BlEig}
H\psi=\lambda_j(k) \psi.
\ee
The question is whether the eigenfunction $\psi$ (and thus the corresponding periodic function $p$) can be chosen with a nice dependence of $k$. Whenever the multiplicity of the eigenvalue $\lambda_j(k)$ stays constant near a point $k_0$, one can choose the Bloch eigenfunction analytically depending on $k$. See, e.g., \cite{ZaiKreKucPan_umn75,Kuc_floquet} for this standard fact. Whenever the band functions collide and thus the multiplicity changes, one cannot extend the Bloch eigenfunction even continuously w.r.t. $k$. However, cutting along these thin bad sets of quasimomenta, one can establish the following

\bl \emph{(Wilcox \cite{Wil_jam78})} \label{L:piecewiseBloch}
There exists a null-set $\mathcal{Z}\subset \T^*$ and a sequence $\psi_j(k)$ of Bloch eigenfunctions, analytic on $\T^*\setminus \mathcal{Z}$, such that  $\{\psi_j(k)\}$ form an orthonormal basis in $L^2(W)$ for $k\in \T^*\setminus \mathcal{Z}$.
\el

This immediately leads to the eigenfunction expansion of any function $f(x)\in L^2(\R^n)$ into Bloch eigenfunctions:
\bt\label{T:Gelfand}(\cite{Gel_dansssr50}, see also \cite{Lidskii_expansions})
Let $f\in L^2(\R^n)$ and
\be\label{E:gelfand0}
\U f(\cdot,k)= \sum_j f_j(k) \psi_j(\cdot, k)
\ee
be the expansion of its Floquet transform into the basis $\psi_j$. Then
\be\label{E:gelfand0}
\|f\|^2_{L^2} = \sum_j \|f_j\|^2_{L^2(\T^*)}.
\ee
\et
In other words,
\bt\label{T:orthExp}
The operator $H$ is unitarily equivalent to the orthogonal direct sum of operators of multiplication by piece-wise analytic continuous scalar functions $\lambda_j(k)$ in $L^2(\T^*)$.
\et

Somewhat weaker results were obtained earlier by Odeh and Keller \cite{OdeKel_jmp64}.

\subsection{Absolutely continuous, pure point, and singular continuous parts of the spectrum}\label{SS:ac,pp,sing}
According to Theorem \ref{T:orthExp}, one needs to understand the spectrum of the operator $A$ acting in $L^2(\T^*)$ as multiplication by a continuous piece-wise analytic scalar function $a(k)$
\be\label{E:MultOp}
(Af)(k)=a(k)f(k).
\ee
\bt\label{T:SpectMult}
The following holds:
\begin{enumerate}
\item The singular continuous spectrum of the operator $A$ is empty:
\be\label{E:NoSc}
\sigma_{sc}(A)=\emptyset.
\ee
\item The pure point spectrum consists of all values $\lambda$ such that the $\lambda$-level set of $a(k)$ has positive measure. In particular, if there is no positive measure ``flat piece'' $a=const$, then $\sigma_{pp}(A)=\emptyset$ and thus the whole spectrum is absolutely continuous:
    $$
    \sigma(A)=\sigma_{ac}(A).
    $$
\item For each $\lambda\in\sigma_{pp}(A)$, the function $a(k)$ is constant on an open set.
\end{enumerate}
\et
We thus get the following result:
\bc\label{C:Nosing}
For any periodic self-adjoint elliptic operator $H$, one has:
\begin{enumerate}
\item The singular continuous spectrum is empty: $\sigma_{sc}(A)=\emptyset$.
\item The pure point spectrum consists of all values $\lambda$ such that the $\lambda$-level set of some of $\lambda_j(k)$ has positive measure.
\item If the pure point spectrum is non-empty, then at least one of the band functions $\lambda_j$ is constant on an open set.
\item The pure point spectrum consists of all values $\lambda_0$, such the complex Bloch variety $B_{H,\C}$ contains the flat component $\lambda=\lambda_0$.
\end{enumerate}
\ec
The last statement of this theorem follows from analyticity of the Bloch variety (Theorem \ref{T:analBloch}), which forces any small piece of a constant branch to extend to the whole flat component. Since Theorem \ref{T:Noflat} prohibits such a situation for the Schr\"odinger operator $-\Delta+V(x)$, one reaches the following conclusion:
\bt\emph{(Thomas \cite{Tho_cmp73})}\label{T:Thomas}
Under the conditions of Theorem \ref{T:Noflat}, the spectrum of the Schr\"odinger operator with periodic potential is absolutely continuous.
\et
As it was the case with Theorem  \ref{T:Noflat}, here the condition of boundedness of the periodic potential is a significantly stronger than needed (see, e.g., \cite[Section VIII.16]{ReedSimon_v14}). One, however, should not get over-excited about dropping conditions on the coefficients in the highest order terms, since when they are going below those guaranteeing weak uniqueness of continuation property, one can get point spectrum, as it was shown by Filonov \cite{Filonov_example}. We will return to this discussion a little bit further in the text.

Since the famous work \cite{Tho_cmp73} by L.~Thomas, all proofs of the absolute continuity of the spectra of periodic elliptic operators followed the same scheme: proving absence of flat components in the dispersion relation (Bloch variety).
Due to the usually hard technical work of many researchers (Birman \& Suslina, Danilov, Derguzov, Filonov, Friedlander, Klopp, Krupchyk \& Uhlmann, Kuchment \& Levendorskii, Morame, Simon,
Shen, Shterenberg, Sobolev, Thomas, and many others), absolute continuity of the spectrum has been proven for many
(\textbf{but still not all}) second order scalar elliptic periodic operators, as well as periodic Pauli, Dirac, elasticity theory, and in some cases
Maxwell operators, see \cite{HempHerbPerMagn,BirSusDirAC,BirSusElastAC,BirSusDirAC2,KrupchUhlmann,BirSus_aa99,BirShtSus_aa00,Danilov_10,Danilov_9,Danilov_03,DanilovMZ_03,FilSob_04,Karpesh_AC08,KloppMAnn10,Sht_zns00,Sob_inv99,TikhFilon_04,Tho_cmp73,ReedSimon_v14,GerNie_jfa98,GerNie_jmku98,ShenZhao_08,Shen_01,ShenIMRN_01,KucLev_tams02,Kuc_floquet,Fri_cmp02,Danilov_11,Danilov_05,Danilov_00,Danilov_00_2,Danilov_99,Danilov_95,Danilov_90} and references therein. However, the following conjecture still remains unproven, even under assumption of infinite differentiability of the coefficients of the operator:

\bcon\label{Co:abscont}
The spectrum of any self-adjoint second order periodic elliptic scalar operator with ``nice'' (e.g., smooth) coefficients is absolutely continuous.
\econ

In all works, except the Friedlander's paper \cite{Fri_cmp02} and few papers based upon it (e.g., \cite{TikhFilon_04}), the proof goes analogously to the one of Theorem \ref{T:Noflat}: dominating in the Fourier domain the lower order terms by the principal part, which gets increasingly hard in presence of variable coefficients in the first order terms (magnetic potential) and it does not look feasible at all to treat variable coefficients in the principal part of the operator. The least technical approach, which beautifully avoids these domination estimates and thus allows for variable coefficients in the second order terms, was due to Friedlander \cite{Fri_cmp02}. Regretfully, it requires some symmetry
condition on the operator, which seems to be superfluous, but no one has succeeded in removing it.

There are also quite a few, often even more demanding, results on absolute continuity for operators in periodic waveguides, on periodic systems of curves and surfaces, etc.
\cite{Sus_rjmp01,BenDucExn_lmp03,ExnerFrank_ahp07,FilKlo_dm04,FilKlo_dm04err,FilKlo_cmp04,Fri_incol04,KachkFilonov_09,KachFil_10,ShtSus_aa01,SobWalth_guides02,Sus_znssp02,SusSht_aa02,Der_vlu72,kuch_floquet1982}.

\br \label{R:4thorder}
Absolute continuity of the spectrum \textbf{fails for some periodic elliptic operators of higher order}.
\er
Indeed, the Plis' example \cite{Pli_jmm60} of a 4th order elliptic operator which fails unique continuation property, i.e. has a compactly supported eigenfunction,
can be easily massaged to produce a periodic example, where $\sigma_{pp}\neq \emptyset$ \cite{Kuc_floquet}.

This relation to the uniqueness of continuation theorems does not seem accidental.
As we have already mentioned, Filonov \cite{Filonov_example} constructed an example of
a 2nd order periodic elliptic operator with non-empty
pure point spectrum, with the leading coefficients falling just below of what is needed for the weak uniqueness of continuation to hold. Another evidence of this is the following result:
\bt\label{T:fastdecay}\cite[Theorems 4.1.5 and 4.1.6]{Kuc_floquet}
Existence of an $L^2$-eigenfunction for a periodic elliptic operator is equivalent to existence of a (different) super-exponentially
decaying eigenfunction, i.e. such that
\be
|u(x)|\leq Ce^{-|x|^\g},
\ee
with some $\g>1$
 (depending on the order of the
operator and dimension, see details in \cite[proof of Theorem 4.1.6]{Kuc_floquet}).
\et
\bcon
Existence of such a solution must violate some ``unique continuity at infinity'' property.
\econ
Regretfully, the only such result known that would apply to periodic operators, due to Froese\&Herbst\& M.\&T. Hoffman-Ostenhof \cite{FroHerHofHof_prse83} and independently
Meshkov \cite{Mesh_decrease}, is not strong enough to lead to new absolute continuity theorems. E.g., it does not allow variable coefficients in the terms of the first and second orders.

Another confirmation of the relation with the uniqueness of continuation comes from equations on discrete or quantum graphs, where the uniqueness of continuation does not hold. And sure enough, one
can find compactly supported eigenfunctions for 2nd order elliptic periodic operators on such graphs (see \cite[and references therein]{BerKuc_book} and Fig. \ref{F:compactlysupported}).
\begin{figure}[ht!]
\begin{center}
  \includegraphics[scale=0.5]{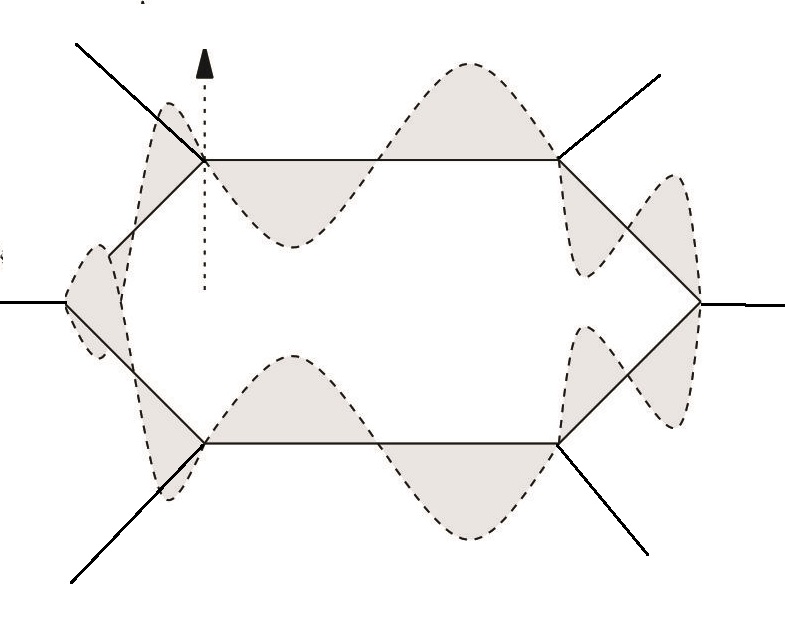}
  \caption{A compactly supported eigenfunction on a quantum graph.}\label{F:compactlysupported}
\end{center}
\end{figure}

\subsubsection{Density of states}\label{SS:density}
An important object in spectral theory and solid state physics is the so called \emph{density of states} of a self-adjoint periodic elliptic operator $H$. Here is the idea: consider an ``appropriate'' sequence of expanding when $N\to\infty$  domains $V_N\subset\R^n$, eventually covering the whole space. After imposing ``appropriate'' boundary conditions on these domains, one can consider the counting function of eigenvalues $\lambda_j$ of the resulting operator with discrete spectrum:
\be\label{E:counting_N}
\#\{\lambda_j<\lambda\}.
\ee
Now one normalizes this function by the volume of the domain $V_N$ and takes the limit when $N\to\infty$:
\bd\label{D:intdensity}
The \textbf{integrated density of states (IDS)} of the operator $H$ is
\be\label{E:IDS}
\rho(\lambda):=\lim_{N\to\infty}\frac{1}{|V_N|}\#\{\lambda_j<\lambda\}.
\ee
\ed

One certainly wonders whether the limit exists and is independent of the choice of the expanding domains $V_N$ and boundary conditions imposed. Under appropriate conditions on those, the answer is a ``yes'' to both questions (see Shubin \cite{Shubin_ap}, where this is done even for almost periodic coefficients, and Adachi and Sunada \cite{AdaSun_cmh93} for a more general discussion).

Without listing the conditions, we just acknowledge that, assuming that the lattice is transformed to $\Z^n$, this is true when one picks $V_N$ as the sequence of cubes $\{x\in\R^n\,| |x_j|\leq N,\, j=1,\dots,n\}$ for $N\in\Z^+$ and one imposes periodic boundary conditions on those (recall our previous discussion of Born-Karman conditions).

Then a simple calculation shows that the density of states can be described in terms of the dispersion relation $\lambda_j(k)$ as follows:
\bt\label{T:IDS}
\be\label{E:IDS_per}
\rho(\lambda)=\sum_j\mu\{k\in\T^*\, |\,\lambda_j(k)<\lambda\},
\ee
where $\mu$ is the previously defined Haar measure on the torus $\T^*$.
\et
Due to the analytic nature of the band functions we have studied, one gets
\bc\label{C:IDSanal}\indent
\begin{enumerate}
\item The IDS is a piecewise analytic function of $\lambda$.
\item Unless a flat band function $\lambda=const$ is present, the IDS is continuous.
\item Singularities of the IDS can arise if $\lambda$ is a singular or critical value of the dispersion relation (\textbf{Van Hove singularities}).
\end{enumerate}
\ec
\bd\label{D:DS}
The \textbf{density of states (DS)} is the Radon-Nikodim derivative of the IDS with respect to the Lebesque measure:
\be\label{E:DoDS}
g(\lambda):=\frac{d\rho}{d\lambda}.
\ee
\ed
\bt\label{T:DS}
\be\label{E:DS}
g(\lambda)=(2\pi)^{-n}\sum_j\int_{\lambda_j=\lambda}\frac{ds}{|\nabla_k \lambda_j|}.
\ee
\et
The Figure \ref{F:IDS} illustrates these notions\footnote{The picture is borrowed from http://www.nature.com/srep/2014/140721/srep05709/images\_article/srep05709-f1.jpg}:
\bef[ht!]
\begin{center}
  \includegraphics[scale=0.5]{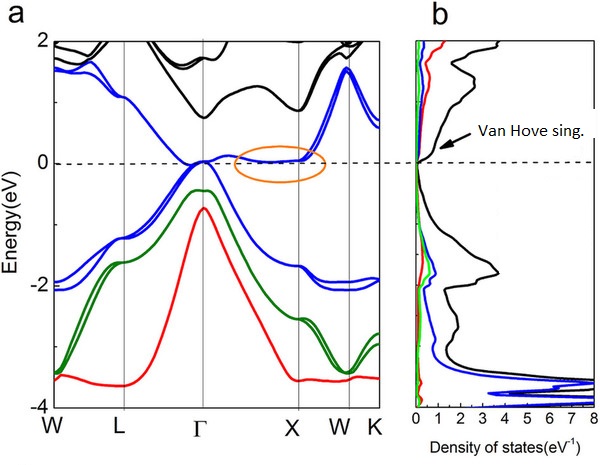}
  \caption{a) Dispersion relation. b) The corresponding density of states.}
  \label{F:IDS}
\end{center}
\eef
Noticing that in (\ref{E:DS}) one encounters the integral of a holomorphic form over a real cycle on the Fermi surface $F_\lambda$, allows one to use powerful techniques of several complex variables (see Gerard \cite{Ger_bsmf90} and Gieseker\&Kn\"orrer\&Trubowitz \cite{GieKnoTru_cm91,Gieseker_fermi}), which we are unable to address here.

\subsection{Spectral asymptotics}\label{SS:asympt}

Another important and active area of research that I cannot discuss in any details, is high energy spectral asymptotics (e.g., asymptotic of density of states). The reader is referred to the following (incomplete) literature list and references provided there: in $1D$ \cite{SobDens1D,ShubDens1D,SobAsy1D,Kappeler_asymptotics}; in higher dimensions \cite{LevendMagnFloAsymp,LevendSpAsympt,KorPushn,SjoMicroloc,Fri_cpde02,ParnShter09,MohElectromagn,Bratelli,HelfMochAsympt,KarpDens00,MorParPchLowerboundDensity,MParnStCompleteExp,ParnStComplExp09,ParnStComplExp12,SobIDS05}, with the paper \cite{ParnStComplExp12} by Parnovski and Shterenberg being probably the most advanced and up to date.

\subsection{Wannier functions}\label{SS:wannier}

Bloch functions clearly are analogs for the periodic case of plane waves $e^{i\xi\cdot x}$, which are localized in Fourier domain. Standard Fourier transform converts the plane waves into delta functions, which are localized in the physical space. One wonders whether there is an analog of the delta function basis for the periodic situation, and if yes, whether such functions are useful. The answer is a ``yes'' for both questions, although new non-trivial issues arise here.

The functions in question are called \textbf{Wannier functions}, which are used frequently in numerical computations in solid state physics and photonic crystal theory \cite{BusFreLin_pre06,BuschWannier,MarzSou}, since they lead to nice discretizations, close to the tight-binding models.
One expects (in analogy with the plane waves and delta functions) that they could arise as inverse Floquet transforms w.r.t. the crystal momentum, of Bloch functions.

Let us try to be more precise.

\bd\label{D:wannier}
Suppose that we have a Bloch function $\phi_k(x)$ (see (\ref{E:BlEig})) that depends ``sufficiently nicely'' on the quasi-momentum $k$. The corresponding \textbf{Wannier function} is defined as follows:
\be\label{E:wannierf}
w(x):=\int_{\T^*}\phi_k(x)dk.
\ee
\ed
Standard Fourier series argument shows that \textbf{smoothness of $\phi_k$ with respect to $k$ translates into decay of $w(x)$}. For instance, infinite differentiability implies that the $L^2$-norm of $w(x)$ on a cube decays faster than algebraically with the shift of the cube to infinity. Analogously, analyticity of $\phi_k$ would produce exponential decay of the Wannier function.

Suppose now that one has a multiple band $S$ consisting of $m$ bands of the spectrum, which is separated from the rest of the spectrum by gaps. As it was discussed in Section \ref{SS:BlochBundle}, one obtains an $m$-dimensional analytic Bloch bundle over $\T^*$. If  this bundle is analytically trivial, it has an analytic basis of Bloch functions: $\phi_{k,1}, .... ,\phi_{k,m}$. It is a simple exercise to see (e.g., \cite{KuchWannier}) that the corresponding exponentially decaying Wannier functions $w_j=\int_{\T^*}\phi_{k,j}dk$ and all their $\G$-shifts form a basis (which can also be made orthonormal under appropriate choice of Bloch functions) in the spectral subspace of the operator $H$ that corresponds to the isolated part of the spectrum $S$.

This construction does not work, if the Bloch bundle is not analytically trivial. An incarnation of Oka's principle, due to Grauert \cite{Grauert}, shows that the only obstacle is topological: if the Bloch bundle is topologically trivial, then it is automatically analytically trivial (see also \cite{ZaiKreKucPan_umn75} for the related discussions). For a while this issue had not been understood and there was a belief that nice Wannier bases always exist, till Thouless showed \cite{Thouless} that in presence of magnetic terms in a periodic Schr\"odinger operator, the Bloch bundle can be non-trivial (such non-triviality also arises in topological insulators \cite{BernTopIns}). Many efforts have been concentrated on establishing sufficient conditions of the triviality, as well as on efficient finding the Wannier bases (see, e.g., \cite[and references therein]{PanatiBlochBundle,KuchWannier,PanatiWann,Marzari,MarzSou,Nenciu}).

In the presence of the topological obstacle, however, no basis of (even slowly decaying) Wannier functions exists \cite{KuchWannier}. However, it was shown that instead of the (non-existing) analytic basis of the bundle, one can always find a ``nice'' Parseval frame (i.e., a ``nice'' overdetermined system) of exponentially decaying Wannier functions \cite{KuchWannier}. Rather crude estimates on the number of extra functions needed is also provided in  \cite{KuchWannier}. However, in physical dimensions, instead of $m$ families of Wannier functions, one only needs $(m+1)$  of these \cite{KuchAu}.

\subsection{Impurity spectra}\label{SS:Imp}

Let us return now to the periodic Schr\"odinger operator $H=-\Delta +V(x)$. An important and frequently studied issue is the existence and properties of the \textbf{impurity spectrum} arising when a localized (compactly supported or sufficiently fast decaying) perturbation $W(x)$ is added to the potential. The well known theorem claims that only eigenvalues of finite multiplicity can appear, leaving the otherwise absolutely continuous spectrum unchanged (see, e.g. \cite[Section 18 and references therein]{Glazman_book}).

Since $V$ is periodic and thus has band-gap structure of the spectrum, these eigenvalues have two options: to appear in the spectral gaps, or embed into the AC spectrum (\textbf{embedded eigenvalues}), see Fig.\ref{F:embed}.
\bef[ht!]
\begin{center}
  \includegraphics[scale=0.7]{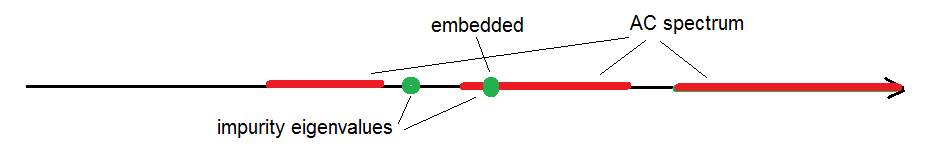}\\ \caption{Impurity eigenvalues on a band-gap structured spectrum.}\label{F:embed}
\end{center}
\eef
The general wisdom is that when the perturbation decays sufficiently fast, embedded eigenvalues cannot arise, while if the decay is not fast enough, one can indeed have them. In the case when the background potential $V$ is equal to zero, starting with the famous work by von Neumann and Wigner \cite{VonNWign}, many results of this nature have been obtained, see e.g. a nice survey in \cite{EasthamKalf_contspec}.

If, however, the background periodic potential is present, the issue is far from being resolved. The results confirming the ``general wisdom'' were obtained in the ODE case by Rofe-Beketov \cite{Rof_dansssr63,Rof_incol84,Rof_mpfa73,Rof_smd64}.

Although similar answers are expected in the higher dimensions, proving them happens to be difficult. Some results were obtained by Vainberg and the author
\cite{KucVai_cpde00,KucVai_incol98,KucVai_cmp06}. E.g.,
\bt\label{T:KuVa}\cite{KucVai_cpde00,KucVai_incol98}
Let $n\leq 4$ and the perturbation $W(x)$ decay exponentially. If for a given $\lambda\in\sigma(H)$
the Fermi surface at that level is irreducible (modulo periodicity), then $\lambda\notin \sigma_{pp}(H+W)$.\\
In particular, if all Fermi surfaces are irreducible, no embedded eigenvalues can arise.
\et
\br\label{R:nodependenceonW}
Notice that irreducibility of the Fermi surface at a level does not depend on the impurity potential $W$.
\er
As our previous discussion has indicated, it is hard to establish irreducibility of the Fermi surface, although for a second order periodic ODE it is automatic. One wonders whether the irreducibility condition of this theorem is just an artifact of the techniques used in the proof.
It is possible, but does not seem to be very likely. The proof indicated that reducibility could in principle lead to a situation similar to the waveguide theory, where impurity eigenvalues can get embedded. And indeed, there are examples of a periodic fourth order ODE \cite{Papanic_emb} (V.~Papanicolaou) and of second order multi-periodic quantum graph operators \cite{shipman_qgr} (Shipman), where reducible Fermi surfaces arise and impurity eigenvalues can be embedded.

In the graph case, due to non-trivial topology, embedded eigenvalues can and do arise. However,
\bt\label{T:KuVa_gr}\cite{KucVai_cmp06} In the graph case, embedded eigenvalues might arise, but if the relevant Fermi surface is irreducible, the corresponding eigenfunction must be
supported ``close'' to the support of the perturbation. (See details in \cite{KucVai_cmp06}.)
\et
It is interesting to note that in the reducible case, eigenfunctions can spread and indeed have unbounded support (Shipman \cite{shipman_qgr}).

\section{Threshold effects}\label{S:Threshold}

The name \textbf{threshold effects} (as coined by Birman and Suslina), is used here for the features that depend upon the infinitesimal structure, or maybe just on a finite \textbf{jet}\footnote{I.e., truncated Taylor polynomial, see https://en.wikipedia.org/wiki/Jet\_(mathematics).}, of the dispersion at a spectral edge. A popular nowadays example is of \textbf{graphene}, where the Dirac cones lead to conductance as if governed by the massless Dirac equation \cite{KatsGraph,FefWei_pre12,FefWeiPackets}. We will not dwell on this particular example, but rather list a number of other important relations.

Regretfully, in order to keep the (already excessive) length of the text in check, the author can afford only brief pointers rather than any detailed discussion or formulation of the results.

\subsection{Homogenization}\label{SS:homog}

Homogenization is an effective medium theory for (long) waves in a highly oscillating periodic (or even random) medium. It is probably the best known threshold effect. Indeed, finding the homogenized (effective medium) operator is, roughly, equivalent to determining the second order jet of the dispersion relation at the bottom of the spectrum (see, e.g. \cite{BirSus_aa03,BirSus_otaa01,AllaireConcaBloch,Bensoussan_periodic,Jikov_homogenization,Conca_FourierHomog,Conca02_BlochApproxHomog,Conca05_BlochHomogBDDDomain,ConcaAllair98_homogBloch,Conca97_homogBloch} and references therein). It is sometimes a highly non-trivial procedure (e.g., for Maxwell operator).

While the usual homogenization leads to a homogeneous medium (which gave the name to the area), and so cannot address finite spectral gaps, more sophisticated approaches (see, e.g. \cite{Zhikov_gaps,HemLie_cpde00,Fri_cpde02}) can achieve this and more.

It is natural to ask whether there is a version of homogenization that occurs near the spectral edges of the internal gaps, where the parabolic shape of the dispersion relation might resemble the one at the bottom of the spectrum. This indeed happens to be the case, as it was shown in a series of works by Birman and Suslina \cite{Bir_aa03,BirSus_zns04}, see also Suslina and Kharin\cite{SusKha09,SusKha11}.

\subsection{Liouville theorems}
The classical Liouville theorem says:
\bt\label{T:Liouville}
Any harmonic function in $\R^n$ of polynomial growth of order $N$ is a polynomial of degree $N$. The dimension of this space is
\be\label{E:HarmDim}
\left(
                                                  \begin{array}{c}
                                                   n+N \\
                                                    N \\
                                                  \end{array}
                                                \right) -  \left(\begin{array}{c}
                                                   n+N-2 \\
                                                   N-2 \\
                                                  \end{array}
                                                \right).
\ee
\et
S.~T.~Yau posed a problem \cite{Yau_cpam75} of generalizing this theorem to harmonic functions on noncompact manifolds of nonnegative curvature, which was resolved by Colding and Minocozzi \cite{ColMin_am97} (there were also partial contributions from various researchers, in particular P. Li, see the references in \cite{Li_incol00,Li_book}). Thus, finite-dimensionality of the spaces of harmonic functions of a given polynomial growth, as well as estimates (rather than exact formulas) for their dimensions were obtained.

On the other hand, a wonderful observation was made by Avellaneda and Lin \cite{AveLin_crasp89} and Moser and Struwe \cite{MosStr_bsbm92}: the spaces of polynomially growing solutions of periodic divergence type second order elliptic equations in $\R^n$ are finite-dimensional. Moreover, their dimensions are given by (\ref{E:HarmDim}). Homogenization techniques were used in both cases, which restricted the consideration to the bottom of the spectrum only. In Pinchover and author's works \cite{KucPin_jfa01,KucPin_tams07} a wide range generalization of this result was obtained: for an elliptic periodic operator on an abelian covering of a compact manifold (or graph) it was shown that validity of Liouville theorem at some energy level $\lambda$ is equivalent to the corresponding Fermi surface consisting of finitely many points (i.e., essentially $\lambda$ being at a spectral edge). Moreover, dimensions of the spaces of polynomially growing solutions were explicitly computed in terms of the lowest order of a non-trivial Taylor expansion term of the dispersion equation at the corresponding spectral edge.

\subsection{Green's function}
Another threshold effect is the behavior of the Green's function (the Schwartz kernel of the resolvent) near and at spectral edges. There are well known general resolvent exponential decay estimates with the exponential rate depending upon the distance to the spectrum, e.g. the Combes-Thomas estimates \cite{CombThom,BarComHis_hpa97}. However, being applied to self-adjoint periodic elliptic operators, these estimates are not very precise. Indeed, first of all, one expects the exponential decay to be direction-dependent, while the operator estimates mentioned above would provide only an isotropic estimate. Secondly, the off-the-spectrum exponential decay must be accompanied by an additional power decaying factor, which the abstract theorems do not provide. On the other hand, at the bottom of the spectrum the dispersion relation of periodic non-magnetic Schr\"odinger operator has, as it has been mentioned before, a non-degenerate (parabolic) extremum \cite{KirSim_jfa87}, i.e. resembles the one for the Laplace operator. Thus, one can hope that at least near the spectrum the decay should resemble the one for the Laplacian case, i.e. involving an additional algebraically decaying factor. Moreover, at the edge of the spectrum one expects (in dimensions three and higher) some algebraic decay.
And indeed, principal terms were found for the asymptotics of the Green's functions of such operators below the spectrum Babillot and by Murata \& Tcushida \cite{MT,Bab_pa98,Bab_aihp88} (see also a simplified derivation for the discrete case at the end of the Woess' book \cite{woess_rwgraphs}). These results confirm the expectation.

One can ask the same question near and at the edges of the internal gaps of the spectrum, as long as the dispersion relation has a non-degenerate (parabolic) extremum there. Results of this type were obtained in the recent works \cite{KuchRaich,KhaKuchRaich,Kha} by Minh Kha, Raich, and the author.

Going inside the spectral bands, it is also natural to mention here the results on the limiting absorption principle for periodic elliptic operators, see e.g. \cite{GerNie_jfa98,Ger_bsmf90,MT06,Gerard_limitabs,BirSusLimAbs}.

\subsection{Impurity spectra in gaps}\label{SS:BirmImpur}

The local structure of the dispersion relation at a spectral gap edge is also important for understanding of appearance and counting of the impurity eigenvalues arising due to a localized perturbation of the potential. See Birman's papers \cite{Bir_fap91,Bir_mt95,Bir_aa96,Bir_aa97} and references therein.

\section{Solutions}\label{S:solutions}
In this section we provide a very brief overview, with pointers to the literature, of various
results concerning solutions of homogeneous and inhomogeneous periodic elliptic equations.

\subsection{Floquet-Bloch expansions}\label{SS:FBexpansion}

The Euler's theorem \cite{Euler}, mentioned in Section \ref{SS:Euler} claims that all solutions of homogeneous constant coefficient linear ODEs (or systems of those) are linear combinations of exponential-polynomial solutions. This classical theorem has an extremely non-trivial generalization to the case of constant coefficient PDEs (sometimes called \textbf{Ehrenpreis' Fundamental Principle}) due to works by Ehrenpreis, Malgrange, and Palamodov, see the books \cite{Leon_fourier,Palam_engl,Palam_Russian} devoted to its proof and applications. Here instead of finite linear combinations of exponential-polynomial solutions, one needs to involve the integrals over the characteristic variety of the operator, which in turn relies upon commutative algebra, algebraic geometry, and several complex variables techniques. One can ask the natural question whether there is an analog of this result for periodic elliptic PDEs, providing expansions into Bloch-Floquet solutions. If yes, one expects it to be harder to prove, due to the non-algebraic nature of the situation (i.e., of the Fermi surface, which replaces the characteristic variety). Nevertheless, somewhat more restrictive than in the constant coefficient case, such results have been obtained in \cite{Kuc_floquet,Kuc_rms82,KuchZel78,Palam_periodic,Kuch_molch}. Most of the book \cite{Kuc_floquet} is devoted to their proofs. Regretfully, probably due to a more restricted nature of these results, they do not seem to have much of consequence. Thus, we do not address them in any detail here.

\subsection{Generalized eigenfunctions and Shnol'-Bloch theorems}\label{SS:shnol}
One of the consequences of the Floquet theory is that for periodic elliptic operators $H$ detecting whether $\lambda$ is in the spectrum is equivalent to existence of a non-trivial Bloch solution $u$ with a real quasi-momentum $k$ of the equation $Hu=\lambda u$. Since the solution is not square integrable, we would call it a \textbf{generalized eigenfunction}. Since one can easily come up with a generalized eigenfunction (just an exponent) of Laplace operator for arbitrary $\lambda\in\C$, it is clear that existence of a growing generalized eigenfunction does not mean being on the spectrum. However, Bloch solutions with real quasimomenta are bounded, one can ask whether presence of a bounded generalized eigenfunction detects the spectrum. This is indeed the case and the following \textbf{Bloch theorem} (probably never proven by Bloch) holds:
\bt\label{T:Bloch}\cite[Section 4.3]{Kuc_floquet}
Existence of a non-trivial bounded solution of a periodic elliptic equation $Hu=\lambda u$ implies existence of a Bloch solution with a real quasi-momentum, and thus $\lambda\in\sigma(H)$.
\et
Indeed, a stronger type of result is known, the so called \textbf{Shnol' Theorem} \cite{Shn_ms57,Shu_sedp90,Kuc_floquet,Glazman_book,Cycon_schrodinger}, which holds also in non-periodic case. Its simplest version is:
\bt\label{T:Shnol}
If for any $\epsilon>0$ there exists a non-trivial generalized eigenfunction $u_\epsilon$ with the estimate
\be
|u_\epsilon (x)|\leq e^{\epsilon|x|},
\ee
then  $\lambda\in\sigma(H)$.
\et
Moreover, the following stronger version holds:
\bt\label{T:Shnol2}
If for some $a>0$ there exists a non-trivial generalized eigenfunction $u$ with the estimate
\be
|u (x)|\leq e^{a|x|},
\ee
then  $dist\left(\lambda,\sigma(H)\right)\leq C\sqrt{e^{2a}-1}$, where $C$ depends only on the operator $H$.
\et
The detailed (although somewhat outdated) discussion of the Shnol' type can be found in \cite[Section 54]{Glazman_book}.

In the periodic elliptic case one can formulate a stronger version \cite[Theorem 4.3.1]{Kuc_floquet}:
\bt\label{T:Shnol}
Let $L(x,D)$ be a scalar periodic elliptic operator with smooth
coefficients in $\R^n$. If $Lu=0$ has a non-zero solution satisfying
the inequality
\be\label{E:expgrowth}
|f(x)| \leq Ce^{a\sum|x_j|}
\ee
for some $a>o$, then it also has a Bloch solution with a quasimomentum $k$ such that
$\Im k_j\leq a$ for all $j=l, ... , n$, that is, a Bloch solution that also satisfies (\ref{E:expgrowth}).
\et
In other words, there is a complex quasi-momentum vector, whose distance to the unit torus (and thus $dist\left(\lambda,\sigma(H)\right)$) can be estimated from above.
\br\label{R:shnol}
The Shnol' theorem stated with the point-wise estimates like in (\ref{E:expgrowth}) fails in non-euclidean situation. E.g., the hyperbolic Laplacian has a bounded generalized eigenfunction for $\lambda=0$, although $0$ is not in the spectrum. Analogous situation occurs for operator on trees. However, analysis of the proof (e.g., in \cite{Glazman_book}) shows that in fact only an integrated ($L^2$) estimate of growth is used, which implicitly incorporates the rate of the ball volume growth. Thus, there is an $L^2$-version that holds even when the volume of the ball grows exponentially (see \cite[Section 3.2]{BerKuc_book} for the graph case).
\er

For further discussions of generalized eigenfunction expansions the reader can refer to \cite{Berezanskii_expansions,ShubinBerezin_schrod,KleinGeneralized} and references therein.

\subsection{Positive solutions}
It was shown by Agmon \cite{Agmon_positive} that positive solutions of a periodic elliptic second order equation in $\R^n$ (or on a co-compact abelian covering) allow integral expansions into positive Bloch solutions. See also \cite[Section 4.6]{Kuc_floquet} for the description of this result, as well as Lin and Pinchover \cite{LinPinchover_manifolds} for the case of nilpotent co-compact covering. See also \cite{Pinch_88Duke,Pinsky} for related discussions.

\subsection{Inhomogeneous equations}

It is well known in the theory of periodic ODEs (e.g., \cite{Arscott,Iakubovich_periodic}) that unique solvability of inhomogeneous equations in $L^2$ or in the space of bounded functions is equivalent to absence of Floquet multipliers of absolute value one (equivalently, absence of Bloch solutions with real quasimomenta). A simple PDE analog of this result (as well as solvability in the spaces of exponentially decaying functions) also holds \cite[Section 4.2]{Kuc_floquet}.

\section{Miscellany}

\subsection{Parabolic time-periodic equations}\label{SS:evolution}

Fluid dynamics problems of stability of periodic flows (Yudovich \cite{Yudovich_book,Yudovich_lect}) lead naturally to the task of developing an analog of Floquet theory (e.g., completeness of and expansion into Floquet solutions) for parabolic time periodic problems in Banach and Hilbert spaces:
\be\label{E:parab}
\frac{dx}{dt}=A(t)x, x(0)=x_0, A(t+1)=A(t).
\ee
The simplest, albeit already important and non-trivial example is the heat equation
\be\label{E:heat}
 \frac{du}{dt}=\Delta_xu+b(x,t)u
\ee
with time periodic function $b(x,t)$ in an infinite cylinder, with Dirichlet or Neumann boundary conditions.
Embarrassingly, there is no general result guaranteeing existence of at least one Floquet solution of (\ref{E:heat}), even less completeness of such solutions. Forget the Lyaponov theorem!

Only in one spatial dimension an impressive result of this kind for (\ref{E:heat}) is achieved in beautiful works by Chow, Lu, and Mallet-Paret
\cite{Chow_94,Chow_95}, where inverse scattering method is used. Nothing comparable is available in higher dimensions, unless severe restrictions are imposed on the periodic term \cite[Ch. 5 and references therein]{Kuc_floquet}. What seems to be the problem? After all, these equations are hypoelliptic, and the general Floquet theory techniques to a large extent applies to such equations \cite[Ch. 3]{Kuc_floquet}!
The thing is that the general operator theory approach to parabolic periodic equations does not work nearly as nicely as it does for the elliptic case. Namely, as it was discovered by Miloslavski\'i (see \cite{Milosl_diss,Milosl1,Milosl2,Milosl3,Milosl4,Milosl5,Milosl6,Milosl7} as well as \cite[Ch. 5]{Kuc_floquet} for the results and discussion), there are extremely nice abstract periodic equations (\ref{E:parab}) with constant highest order terms, which have no Floquet solutions at all. The positive results known  (\cite{Milosl_diss,Milosl1,Milosl2,Milosl3,Milosl4,Milosl5,Milosl6,Milosl7} and \cite[Ch. 5 and references therein]{Kuc_floquet}) require very strong restrictions on periodic terms, such as $b(x,t)$ in (\ref{E:heat}). The author hopes that the ``pathological'' non-existence of Floquet solutions does not hold for (\ref{E:heat}) (all known counterexamples are abstract parabolic equations). The similar difficulty is also encountered in the Floquet theory for evolution equations (\ref{E:parab}) with a bounded operator coefficient $A(t)$ (see \cite{DalKrein}).

Some results on positive solutions of periodic parabolic equations have been obtained (see, e.g. \cite{PinchPosParab,Ghor_perparab,Hess_perparab}).

Time periodic \textbf{hyperbolic and Schr\"odinger type equations} have also been attracting a lot of attention in physics (e.g., \cite{Zeld,Gavrila,Chu}).
However, most of the analytic theory described before fails here and other techniques are required and are being developed. See, e.g. the survey by Yajima \cite{YajimaSurv} and references therein, as well as, e.g. \cite{SviridovComplet,SviridovDiss,YajimaLargeTime,YajimaScat,Howland_79,Howland_89,Howland_92,Howland_98}.

\subsection{Semi-crystals}\label{SS:semi}
So far, we have considered infinite periodic structures. The case of semi-crystal, where a periodic medium occupies a half-space, with another (homogeneous or periodic) medium taking the rest of the space, is extremely important. Here one is interested in scattering on and transmission through the semi-crystal, existence of surface (edge) states, etc. One is referred, for instance, to Ch. 5 in Karpeshina's book \cite{Karp_LNM} and references therein (e.g., to Chuburin's work). See also somewhat related works \cite{Fra_dm03,FraSht_dm04,FefWeinEdgeHoney,DavSim,SmyshKuch}. It seems to the author that this topic has not been sufficiently studied yet.
\subsection{Photonic crystals}\label{SS:PBG}
Photonic crystals are artificial periodic optical media that are optical analogs of semiconductors, which bring about major technological advances (see
\cite{Kuch_pbg,PBGbibl,Joannopoulos_photocrystals} for a nice physics introduction and bibliography, as well as \cite{Kuch_pbg,PBG_Dorfler} for mathematics surveys).
The main equation to study here is Maxwell operator with periodic coefficients. Floquet theory for this operator works in many regards in parallel with the elliptic equations considered in this article, with some notable analytic and numerical quirks (see, e.g. \cite{FigKuc_siamjam96,FigKuc_siamjam96a,FigKuc_siamjam98,Kuch_pbg,Morame_acMaxwell} for details and references).

\subsection{Waveguides}\label{SS:guides}
Periodic waveguides form one of the important applications of the Floquet theory. Here one considers boundary value problems in periodically shaped domains, where the boundary conditions and the governing equations are also periodic. In some applications, such as quantum waveguides or photonic crystal waveguides, one has to deal also with problems on zero-width surface or curve systems, or where the waves rather than being confined to the interior of a waveguide, leak (being evanescent) into the surrounding space. Many spectral problems (e.g., absolute continuity of the spectrum or Bethe-Sommerfeld conjecture) happen to be significantly harder in the waveguide situations and are not completely understood, in spite of significant progress achieved \cite{Der_smj80,Der_thesis75,Der_vlu72,SobWalth_guides02,Kuc_floquet,Fri_incol04,ShtSus_aa01,Sht_zns03,Sus_rjmp01,Sus_znssp02,SusSht_aa02,BrHoPlum,HoangRadosz}.

\subsection{Coverings and non-commutative versions}\label{SS:noncommu}
As we have mentioned before, one can consider periodic elliptic operators on a normal co-compact Riemannian\footnote{Or analytic and even discrete.} covering $X \mapsto M$ with a deck group $\G$.
If $\G$ is a finitely generated virtually abelian group, the techniques and results of Floquet-Bloch theory described in this text mostly apply (e.g., \cite{Sun_jfsut90,KucPin_tams07,KucPin_jfa01,Agmon_positive,BruExnGey_jpa03}), with some caveats such as possible appearance of point spectrum.

Getting rid of the commutativity condition has proven to be hard, due to the difficulties of harmonic analysis on $\G$. And indeed, many techniques fail (after all, Floquet transform belongs to the commutative harmonic analysis) and results do change. There is probably no overarching analog of Floquet theory for all co-compact non-abelian coverings.

Some results do translate to the nilpotent case (Lin and Pinchover \cite{LinPinchover_manifolds}), albeit apparently much more should be possible (e.g., some Liouville type theorems).

In the presence of a periodic magnetic field (but not necessarily periodic magnetic potential), the Schr\"odinger operator is not periodic anymore, but it commutes with the so called \textbf{Zak transformations} (\textbf{magnetic translations}), which combine spatial shifts with phase shifts \cite{LevendMagnFloAsymp,Zak,Novikov_viniti,Auslander_nilmanifolds,Auslander_transform}. Here one deals with the action of the (nilpotent) discrete Heisenberg group $\G$. This case has been intensively studied, due to its physics importance (e.g., \cite[and references therein]{Novikov_viniti,Dinaburg,SjoMicroloc}), but we cannot cover it here.

In more general situations, many issues of spectral theory of periodic Laplacians (and more general operators) on co-compact coverings have been studied, such as relations to the amenability of $\G$ \cite{Bro_cmh81}, existence of a band-gap structure \cite{BruSun_nmj92,BruSun_nmj92,Sun_cjm92}, density of states \cite{AdaSun_cmh93,EfrShubHYperb}, Shnol/Bloch theorems \cite{KobOnoSun_fm89}, presence of pure point spectrum  \cite{KobOnoSun_fm89}, gaps creation \cite{Green,GreenPhD,LledPostResid,Kord05,Kord06,MathShub}, absence of singular continuous spectrum \cite{GruberMeasFermi}, representation of solutions \cite{KuchSpSynth,KuchSymmDAN,KuchRepresSym}, etc. See also \cite{BruSun_a92,Bro_cmh81,Sun_lnm88,GruberNoncom,GrubPosMeas} for related considerations.
The reader is reminded that the references in this section are especially far from being  comprehensive.

\section{Acknowledgments}
This work was partially supported by the NSF grant DMS-1517938. The author expresses his deep gratitude to the NSF for the support. The author also would like to thank the Isaac Newton Institute for Mathematical Sciences, Cambridge, for support and hospitality during the programme Periodic and Ergodic Problems, where work on this paper was undertaken. Thanks also go to Prof. S.~Friedlander, without whose encouragement this survey most probably would not have seen the light of the day.

The author is indebted to many colleagues, co-authors, and former and current students for their publications, discussions, and encouragement, which have made possible over the years my work on this subject.
The list of these people is so long, that the author has decided, with apologies to many, to postpone it till the planned monograph. In here, I express gratitude to the referee, as well as to Ngoc~Do, N.~Falkner, N.~Filonov, L.~Friedlander, T.~Kappeler, Yu.~Karpeshina, Minh~Kha, H.~Kn\"orrer, K.~Krupchyk, P.~Milman, B.~Mityagin, Y.~Pinchover, L.~Parnovskii, B.~Simon, R.~Shterenberg, A.~Sobolev, and T.~Suslina for comments, corrections, suggestions, and improvements of pictures. The author is solely responsible for any remaining errors, incompleteness, misnaming the results, etc.


\bibliography{MKperiodic}{}
\bibliographystyle{amsplain}

\end{document}